\documentclass[useAMS,usenatbib]{mn2e}
\usepackage{graphicx}
\usepackage{amsmath}

\usepackage{epsfig}
\usepackage[dvips]{color}

\usepackage{subfigure}
\usepackage{longtable}\usepackage{lscape}
\usepackage{times}
\DeclareGraphicsExtensions{.pdf, .jpg}
\title[Instability of the torus and QPOs]{On the development of the Papaloizou-Pringle 
instability of the black hole-torus systems
and quasi-periodic oscillations}
\author[O. D\"{o}nmez]
{O. D\"{o}nmez$^{1}$\thanks{E-mail: odonmez@nigde.edu.tr (OD)}\\
$^{1}$Nigde University, Department of Physics, Nigde Turkey, 51200}

\begin{document}

\date{Received 2013}

\pagerange{\pageref{firstpage}--\pageref{lastpage}} \pubyear{2013}

\maketitle

\label{firstpage}

\begin{abstract}
We present the numerical study of dynamical instability of a pressure-supported relativistic torus,  
rotating around the black hole with a constant specific angular momentum on a fixed space-time
background, in case of perturbation by a matter coming from the outer boundary. Two dimensional 
hydrodynamical equations are solved at equatorial plane using the 
HRSCS to study the effect of perturbation on the stable systems. We have found that the 
perturbed torus creates an instability which causes the gas falling into the black hole
in a certain dynamical time. All the models indicate an oscillating torus with certain frequency 
around their instant equilibrium. The dynamic of the accreted torus  varies with the size of initial 
stable torus, black hole spin and other variables, such as Mach number, sound speed, cusp 
location of the torus, etc. The mass accretion rate is slightly 
proportional to the torus-to-hole mass ratio in the black hole-torus system, but 
it strongly depends on the cusp location of the torus. 
The cusp located in the equipotential surfaces of the effective
potential moves outwards into the torus. The dynamical change of the torus increases the
mass accretion rate  and  triggers the  Papaloizou-Pringle instability. 
It is also observed that
the growth of the $m=1$ mode of the Papaloizou-Pringle instability occurs for a 
wide range of fluid and hydrodynamical parameters and a black hole spin. 
We have also computed the QPOs from the oscillating relativistic torus.\\
\end{abstract}


%%%%%%%%%%%%%%%%%%%%%%%%%%%%%%%%%%%%%%%%%%%%%%%%%%%%%%%%%%%%%%%%%%%%%%%
%%%%%%%%%%%%%%%%%%%%%%%%%%%%%%%%%%%%%%%%%%%%%%%%%%%%%%%%%%%%%%%%%%%%%%%
%%%%%%%%%%%%%%%%%%%%%%%%%%%%%%%%%%%%%%%%%%%%%%%%%%%%%%%%%%%%%%%%%%%%%%%
%%%%%%%%%%%%%%%%%%%%%%%%%%%%%%%%%%%%%%%%%%%%%%%%%%%%%%%%%%%%%%%%%%%%%%%
%%%%%%%%%%%%%%%%%%%%%%%%%%%%%%%%%%%%%%%%%%%%%%%%%%%%%%%%%%%%%%%%%%%%%%%
%%%%%%%%%%%%%%%%%%%%%%%%%%%%%%%%%%%%%%%%%%%%%%%%%%%%%%%%%%%%%%%%%%%%%%%

\section{Introduction}
\label{Introduction}

 The coalescing of compact binaries and their interactions with accretion discs are important 
issues in astrophysics. The recent review about the black hole accretion disc theory given by \citet{AbramFrag}
explains how the interaction of accretion disc with a black hole reveals predictions of emission, signature 
of strong gravity, black hole mass, and spin.  
The merging of BH-BH, BH-NS or NS-NS, or accreting a hot matter rotating around compact objects
 may cause a torus to be formed which might be responsible for 
oscillation and emission of BH-torus systems. The hot torus around the black hole is thought to be 
a mechanism gather the particles and form a jet. The internal shocks inside the jets can create
high energy particles and lead to the emission of gamma-ray photons\citep{Meszaros1}.  It is important
to understand the dynamics, formation,  and stability properties of a torus to show how the gamma rays are formed.

The rapid destruction of an accretion disc due to high accretion rate, which is larger than the Eddington
rate, was first suggested by \citet{AbrCalNob}. They pointed out that the cusp located in the equipotential 
surfaces of the effective potential moves outwards into 
the torus. This physical change in the black hole-torus system  can increase the mass accretion rate and 
mass of the black hole, and it causes  a runaway instability.
The runaway instability was  firstly discovered by \citep{AbrCalNob}
using a pseudo-Newtonian potential and studied in relativity by  \citet{FonDai}.
The comprehensive study of relativistic thick disk  which has a non-constant angular momentum around the black holes 
has been done by  \citet{DaiFont}. They found that the disk is dramatically stabilized in case  of very small 
values of angular momentum slopes. The non-constant angular momentum disk displays more unstable behavior and it 
causes to grow the mode of the runaway instability more fast.
The instability of accreting torus orbiting around the non-rotating black hole was 
studied by  \citet{MFS},  and their numerical simulations showed that axisymmetric oscillation 
of torus was exhibited without the appearance of the runaway instability. It is also indicated 
that the self gravity of the torus does not play an important role for the occurring of the instability in 
a few dynamical time steps.  
Recently,  \citet{KASSZRO}, for the first time, demonstrated the runaway instability for 
a constant distribution of the specific angular momentum
by using the fully self-consistent general relativistic hydrodynamical evolution. 
They have confirmed that the runaway instability  occurred in a few dynamical times.

The Papaloizou-Pringle instability represents the instability of the torus around the compact objects.
The origin of 
the Papaloizou-Pringle instability is mostly related with the interaction between  the propagation of waves
across the co-rotation radius \citep{PapPri1}.  The waves inside the co-rotation radius carry negative energy while 
the waves outside of this radius carry positive energy. When the waves inside the 
co-rotation radius lose energy to the  waves outside the co-rotation radius, the instability occurs \citep{BlaGlat}.
The Papaloizou-Pringle instability on the torus was studied by a number of  authors in literature.
The number of physical parameters play a dominant role emerging  from the hydrodynamical instabilities occurred as a 
result of interaction between perturbation and the black hole-torus system. Three-dimensional space  parameter of the 
torus size, the torus angular momentum and angular momentum of the black hole was studied by \citet{DeVHaw}. The growth 
of the Papaloizou-Pringle instability was observed for a range of initial configuration of the black hole-torus
system.
The effect of magnetic field on the Papaloizou-Pringle instability in the torus was revealed by
\citet{FuLai} using the various magnetic structures and strengths. They have found that the toroidal magnetic 
fields affect the Papaloizou-Pringle instability. The self-gravity of the relativistic torus orbiting 
around the black hole may be thought as a central
engine of GRBs. So it is important to understand its dynamic, formation and stability properties. The 
instabilities of  the self-gravitating torus were studied by \citet{KiShMonFo, KASSZRO} using the general relativistic 
hydrodynamics code and they showed that the 
Papaloizou-Pringle instability grows for a self-gravitating relativistic torus around the black hole.

%which 
%arise due to the interaction of a global non-axisymmetric perturbation with a black hole-torus system

%\noindent
The interacting matter with a black hole or falling matter into a black hole is a source of 
gravitation radiation. Perturbation of the torus by matter modifies the fluid  around 
the black hole and shock may be formed. The rotating shock can cause a matter to 
have a non-linear behavior and its energy  heats the gas. So the electromagnetic 
flares would be produced and these flayers continue to occur for a long time  
\citep{ScKr}. 
Interacting particles with shock waves in the equatorial
plane produce a continues emission ranging from  Ultra-Viole to X-rays if they freely 
drive away from the black hole-torus system \citep{ReMes}.

%\noindent
It is known that the central region of the black hole-disc system, which produces the Gamma Ray Burst (GRB),
is not significantly bigger than the size of the black hole. So the torus is a possible physical mechanism to 
produce GRB  and  
gravitational waves which are used to identify the nature of 
the inner region that  may be observed from GRB system \citep{KiShMonFo}.
The small perturbations on the torus create a hydrodynamical instability  which is responsible  for
gravitational waves. Its contribution to the gravitational wave background was estimated by \citet{CowPutBur}.
The gravitational wave radiation from the center of GRB, which consists of a black hole and torus, 
was studied by  \citet{lee1}. It was found that the gravitational wave is a natural phenomenon in the 
wobbling system. 

%\noindent
Oscillation properties of the black hole-torus system have been studied for numbers of astrophysical systems.
Investigating the oscillating system can allow us to expose disco-seismic 
classes which are dominant in 
relativistic geometrically thin disc. Some numerical studies revealed the oscillatory modes of the perturbed
torus \citep{ZRF1, RYZ1, MZFR}. Oscillation properties of the pressure-supported torus were also examined by 
 \citet{ScRe}. 
They had found a strong correlation between intrinsic frequency of the torus,
which is also called the relative frequency and the frequency of oscillation computed from data 
that is dumped from the mean wind,
and extrinsic frequency, which is a frequency recorded by an observer moving with the flow, 
shown in observed light curve power spectrum.
The analysis of Quasi-Periodic Oscillations (QPOs) was motivated by 
observed data coming from the  galactic center, low-mass $X-$ ray binaries \citep{VanDer} and high 
frequency phenomena, testing the strong gravity \citep{BuAbKaKL}	 
in the innermost region of accretion disc around the black hole. 

%\noindent
The Observations with X-ray telescopes have uncovered the existing of QPOs
of matter around the black holes \citep{RMEMJBOJ, Strohmayer1}. The QPOs can be used to test the strong gravity 
and the properties of the black hole since they are originated close to the black hole \citep{DZR,AbramFrag}. 
In order to 
understand these physical properties, the perturbed pressure-supported torus is examined in a close-binary 
system.
In this paper, our main goal is to study the instability of a stable torus on the equatorial plane based on
perturbation, which is coming from the  outer boundary of computational domain. In order to do that we solve
hydrodynamical equations on a curved fixed space-time for the non-rotating and rotating black holes at 
the equatorial plane. Thus, we  investigate the effects of perturbation and thermodynamical 
parameters onto the tori's 
instability. 
It is astrophysically possible that a stable black hole-torus system might be perturbed 
by a blob of gas like observed in $Sgr A^*$.
We have perturbed the torus 
in a different way from the those used in \citet{MZFR, ScRe, ZanRez} in which 
it was performed by applying a perturbation to the radial velocity of the torus 
and \citet{ZFRM} in which three different types of perturbations, such as  the perturbation
of the radial velocity, a small variation of the density of the torus and a linear perturbation
of the density and radial velocity, were used.

%\noindent
In this work, we study the effect of perturbation onto the torus-black hole system by solving
the hydrodynamical equations on a fixed background space-time. The paper is organized as 
follows: in section \ref{Formulation of Relativistic Hydrodynamics}, we briefly explain the formulation,
numerical setup, boundary and initial condition and introduce the initial setup of the relativistic torus. 
The numerical results are described in \ref{Numerical Results}. The outcomes of interaction of
the torus-black hole system with matter as a result of
perturbation coming from the outer boundary and QPOs due to oscillation of the torus are presented
Finally, Section \ref{Conclusion}  concludes our findings. Throughout 
the paper, we use geometrized unit, $G = c =1$ and space-time signature $(-,+,+,+)$. 

%%%%%%%%%%%%%%%%%%%%%%%%%%%%%%%%%%%%%%%%%%%%%%%%%%%%%%%%%%%%%%%%%%%%%%%
%%%%%%%%%%%%%%%%%%%%%%%%%%%%%%%%%%%%%%%%%%%%%%%%%%%%%%%%%%%%%%%%%%%%%%%
%%%%%%%%%%%%%%%%%%%%%%%%%%%%%%%%%%%%%%%%%%%%%%%%%%%%%%%%%%%%%%%%%%%%%%%
%%%%%%%%%%%%%%%%%%%%%%%%%%%%%%%%%%%%%%%%%%%%%%%%%%%%%%%%%%%%%%%%%%%%%%%

\section{Equations, Numerical Setups, Boundary and Initial Conditions}
\label{Formulation of Relativistic Hydrodynamics}
%%%%%%%%%%%%%%%%%%%%%%%%%%%%%%%%%%%%%%%%%%%%%%%%%%%%%%%%%%%%%%%%%%%%%%%

\subsection{Equations}
\label{Equation}

We consider a non-self gravitating a torus with an 
equation of state of a perfect fluid  in a hydrostatics equilibrium around the non-rotating 
and rotating black holes. In order to model the 
instability of the torus due to perturbation, we have solved the 
hydrodynamical equations on a curved fixed space-time using the equation of state of a 
perfect fluid. The hydrodynamical equations are

\begin{eqnarray}
\bigtriangledown_{\mu} T^{\mu\nu} =0, \;\;\;  \bigtriangledown_{\mu} J^{\mu}=0,
\label{Eq.1}
\end{eqnarray}

\noindent
where $T^{\mu\nu} = \rho h u^{\mu} u^{\nu}  + P g^{\mu \nu}$ is a stress-energy tensor for a
perfect fluid and $J^{\mu} =\rho u^{\mu}$ is a current density. To preserve and use the conservative properties
of hydrodynamical equations, Eq.\ref{Eq.1} is written in a conservative form using
the $3+1$ formalism, and we have \citep{FonMiSuTo}

\begin{eqnarray}
\frac{\partial {\bf{U}}}{\partial t} + \frac{\partial {\bf{F}}^r}{\partial r} + 
\frac{\partial {\bf{F}}^{\phi}}{\partial \phi}= {\bf{S}}\ ,
\label{hydro1}
\end{eqnarray}

\noindent
where $\bf{U}$ is a conserved variable and, $\bf{F}^r$ and $\bf{F}^{\phi}$ are fluxes in radial 
and angular direction at equatorial plane, respectively. These quantities depend on fluid and thermodynamical 
variables which are written explicitly in detail in \citet{Orhan, FonMiSuTo}. $\bf{S}$ represents the source term
which depends on space-time metric $g^{\mu \nu}$, Christoffel symbol $\Gamma^{\alpha}_{\mu \nu}$, lapse 
function $\alpha$, determinant of three metric $\gamma_{ij}$ and   stress-energy tensor  $T^{\mu\nu}$.
The conserved variables and fluxes also depend on the velocity of fluid and Lorenz factor. The covariant
components of three-velocity $v^i$ can be defined in terms of four-velocity $u^{\mu}$ as 
$v^i = u^i/(\alpha u^t)$  and Lorenz factor $W =\alpha u^0 = (1 - \gamma_{ij}v^iv^j)^{-1/2}$. 
The explicit representation of hydrodynamical equations on a curved fixed background space-time 
and their numerical 
solutions at equatorial plane are given in \citet{Orhan,DenOrh}.

%%%%%%%%%%%%%%%%%%%%%%%%%%%%%%%%%%%%%%%%%%%%%%%%%%%%%%%%%%%%%%%%%%%%%%%

\subsection{Numerical Setups  and Boundary Conditions}
\label{Numerical Setups  and Boundary Conditions}
%%%%%%%%%%%%%%%%%%%%%%%%%%%%%%%%%%%%%%%%%%%%%%%%%%%%%%%%%%%%%%%%%%%%%%%

We have solved the hydrodynamics (GRH) equations at the 
equatorial plane to model the perturbed torus around the black hole. The code used 
in this paper, which carries out the numerical simulation, was explained in detail by \citet{Orhan,DZR,DenOrh},
which gives the solutions of the GRH equations using Marquina method with MUSCL left and right states.  Marquina method 
with MUSCL scheme guarantees higher order accuracy and gives better solution at discontinuities mostly seen 
close to the black hole. The pressure of the
gas is computed by using the standard $\Gamma$ law equation of state for a perfect fluid which defines 
how the pressure changes with the rest-mass density and internal energy of the gas,
 $P = (\Gamma-1)\rho\epsilon$ with $\Gamma = 4/3$.  
After setting up the stable torus  with a constant specific 
angular momentum inside the region, $r_{in} < r < r_{max}$, we have to also define the vacuum, 
the region outside of the   $r_{in} < r < r_{max}$
at where all the variables set to some negligible values. 
We have used the low density  atmosphere, $\rho_{atm} = 10^{-8}\rho_c$, 
outside of the torus in the computational domain. The other variables of atmosphere are 
$p_{atm} = 10^{-8}p_c$, $V^{r}=0.0$ and $V^{\phi}=0.0$. The numerical evolution of 
torus showed that  the dynamic of steady-state torus was unaffected by 
the presence of the given atmosphere.
The Kerr metric in Boyer-Lindquist coordinates is used to set up the black hole
at the center of computational domain using an  
uniformly spaced grid 
in $r$ and $\phi$ directions.
The $r$ goes from $2.8M$ to $200M$ for non-rotating, and   $1.7M$ to $200M$ for rotating black holes
while $\phi$ varies between $0$ and $2\pi$. Typically, we use
$N_r$ $X$ $N_{\phi}$ $=$ $3072$ $X$ $256$ zones in radial and angular directions at the equatorial plane. 

The boundaries must have been correctly  resolved to avoid unwanted oscillations. So we set up an outflow
boundary condition at physical boundaries of computation domain in radial direction. All the variables
are filled with values using zeroth-order extrapolation. While the velocity of matter, $v^r$, must be 
less than zero at close to the black hole, it should be bigger than zero at the outer boundary of 
computation domain. 
The positive radial velocity at the outer boundary lets the gas to fall out from computational domain.
So the unwanted oscillations which may arise at the outer boundary can not propagate into the computational domain.
The periodic boundary condition is used in angular direction.

%%%%%%%%%%%%%%%%%%%%%%%%%%%%%%%%%%%%%%%%%%%%%%%%%%%%%%%%%%%%%%%%%%%%%%%

\subsection{Initial Torus Dynamics}
\label{Torus Dynamics as a Numerical Setup}
%%%%%%%%%%%%%%%%%%%%%%%%%%%%%%%%%%%%%%%%%%%%%%%%%%%%%%%%%%%%%%%%%%%%%%%

 The analytic representation of the non-self-gravitating relativistic torus for a test fluid 
was first examined 
by \citet{AbrJarSik}. They found a sharp cusp for marginally stable accreting disc which was located 
at an equatorial plane.
Later, the torus with an equation of state of a perfect fluid  was 
discussed in detail in \citet{ZanRez,ZRF1,ZFRM,NFZde}, numerically. 
The matter of torus is rotating at a circular 
orbit with non-geodesic flow between $r_{in}$  and $r_{max}$, and the polytropic equation of state, 
$P = K \rho^{\Gamma}$, is used to build an initial torus in hydrodynamical equilibrium with the values of 
variables given in Table \ref{table:Initial Models1}. Using the $\Gamma =4/3$ , it mimics a degenerate 
relativistic electron gas. The significant internal pressure in the torus can balance
the centrifugal and gravitational forces to maintain the system in hydrodynamical equilibrium. 

Some of the astrophysical systems, which have a torus type structure, may be formed as a 
consequence of the merger of two black holes, two neutron stars, their mixed combination or form in the 
core collapse of the massive stars. These phenomena suggest that the torus might be in non-equilibrium 
state after it is formed. In order to understand the physical phenomena of a non-stable black 
hole torus system and perturbation of the torus by a matter which is coming from the outer boundary of 
computational domain, 
we build  the initial conditions for a non-self gravitating perfect-fluid torus orbiting around a black hole
using the formulations given by \citet{ZanRez,ZRF1} that assumes 
to be a non-geodesic motion of the flow . 
We have considered different initial conditions and perturbations 
given in Table \ref{table:Initial Models1}.

\begin{table*}
%\scriptsize 
%\tiny
\footnotesize 
  \caption{Initial models of the perturbed torus around the 
black hole at the equatorial plane. From left to right:
$P$ shows the model name, $a$ is the spin parameter of 
the black hole,  $M_t/M_{BH}$ is the torus-to-hole mass ratio,  
$K(geo)$ is the polytropic constant in geometrized unit,
$\ell_{0}$ is the constant specific angular momentum,
$r_{in}$ and $r_{out}$ are the inner and the outer edges of the torus,
$r_{cusp}(M)$ is the cusp location,
$\rho_c(geo)$ is the density at the center of the torus, $r_{c}(M)$ represents the location at where density is maximum, 
$t_{orb}(M)$ is the orbital period at $r= r_{c}$, 
$C_{s}$  and $\mathcal{M}$ are the speed of sound and the Mach number of perturbation, respectively.
The total time of the simulations varies depending on models. 
 \label{table:Initial Models1}}
\begin{center}
\vspace*{-2ex}
  \begin{tabular}{ccccccccccc|cc}
    \hline 
    \hline 
        &      &             &          &          &               &    &   &    & &    & $\bf{Param.}$ $\bf{of}$ & $\bf{Perturb.}$  \\
 Model  & $\frac{a}{M}$  & $\frac{M_t}{M_{BH}}$ & $K(geo)$ & $\rho_c(geo)$ & $\ell_{0}$  & $r_{in}$ & $r_{out}$ &$r_{cusp}(M)$ & $r_c(M)$ & $t_{orb}(M)$  &$C_s(c)$ & $\mathcal{M}$ \\
\hline
$P_1$   & $0.0$  & $0.1$ & $4.969$ x $10^{-2}$ & $7.329$ x $10^{-5}$ & $5.44$  & $15.0$ & $72.99$ & $2.91$  & $25$    & $785.3$ & $0.001$ & $2$   \\
$P_2$   & $0.0$  & $0.1$ & $4.969$ x $10^{-2}$ & $7.329$ x $10^{-5}$ & $5.44$  & $15.0$ & $72.99$ & $2.91$  & $25$    & $785.3$ & $0.001$ & $2$   \\
$P_3$   & $0.0$  & $0.1$ & $4.969$ x $10^{-2}$ & $1.140$ x $10^{-4}$ & $3.80$  & $4.57$ & $15.889$ & $4.57$  & $8.35$ & $151.6$ & $0.001$ & $2$      \\
$P_4$   & $0.0$  & $0.1$ & $4.969$ x $10^{-2}$ & $1.140$ x $10^{-4}$ & $3.80$  & $4.57$ & $15.889$ & $4.57$  & $8.35$ & $151.6$ & $0.01$  & $2$    \\
$P_{5}$ & $0.0$  & $0.01$& $3.184$ x $10^{-2}$ & $2.863$ x $10^{-6}$ & $3.95$  & $5.49$ & $29.08$ & $4.107$ & $9.97$  & $197.7$ & $0.001$ & $2$    \\ 
$P_6$   & $0.0$  & $1$   & $2.294$ x $10^{-2}$ & $1.153$ x $10^{-3}$ & $3.80$  & $4.57$ & $15.889$ & $4.57$  & $8.35$ & $151.6$ & $0.001$ & $2$    \\  
$P_7$   & $0.0$  & $0.1$ & $3.601$ x $10^{-2}$ & $1.628$ x $10^{-4}$ & $3.784$ & $4.64$ & $14.367$ & $4.64$  & $8.16$ & $146.4$ & $0.001$ & $2$   \\ 
$P_8$   & $0.0$  & $1$   & $2.294$ x $10^{-2}$ & $1.153$ x $10^{-3}$ & $3.80$  & $4.57$ & $15.889$ & $4.57$  & $8.35$ & $151.6$ & $0.001$ & $200$ \\ 
$P_9$   & $0.0$  & $0.1$ & $3.601$ x $10^{-2}$ & $1.628$ x $10^{-4}$ & $3.784$ & $4.64$ & $14.367$ & $4.64$  & $8.16$ & $146.4$ & $0.001$ & $0.1$ \\ 
\hline
$P_{10}$ & $0.9$ & $0.1$ & $ 4.969$ x $10^{-2}$ & $1.140$ x $10^{-4}$ & $2.60$  & $1.78$ & $19.25$ & $1.78$  & $3.40$ & $39.4$  & $--$    & $--$  \\ 
$P_{11}$ & $0.9$ & $0.1$ & $ 4.969$ x $10^{-2}$ & $1.140$ x $10^{-4}$ & $2.60$  & $1.78$ & $19.25$ & $1.78$  & $3.40$ & $39.4$  & $0.001$ & $2$   \\ 
$P_{12}$ & $0.9$ & $0.05$& $ 4.969$ x $10^{-2}$ & $1.999$ x $10^{-5}$ & $3.303$ & $-$    & $21.23$ & $1.41$  & $7.40$ & $126.5$ & $0.001$ & $2$  \\  
\hline 
\hline 
  \end{tabular}
\end{center}
%  \tablenotetext{}{}
%\vskip -0.8truecm
\end{table*}
%

%%%%%%%%%%%%%%%%%%%%%%%%%%%%%%%%%%%%%%%%%%%%%%%%%%%%%%%%%%%%%%%%%%%%%%%

\subsection{Astrophysical Motivation}
\label{Astrophysical Motivation}
%%%%%%%%%%%%%%%%%%%%%%%%%%%%%%%%%%%%%%%%%%%%%%%%%%%%%%%%%%%%%%%%%%%%%%%

The numerical simulations of the astrophysical systems are getting more popular to understand 
the dynamics of the systems  and to explain the physical phenomena in the results obtained from the observational data.
One of the interesting problems 
in astrophysics is the torus around the black hole. The interaction of torus with a black hole may create 
instability as a consequence of perturbation.  It is believed that the central engine for GRBs
is the black hole-torus system. Microquasars also consist of the black holes which are surrounded by the tori \citep{LevBlan}.
The perturbation is one of the mechanism of the evolution triggering the instability in the black hole-torus system to consider the
observational and physical results of the torus and the black hole. Upcoming and lunched ground- and space-base detectors
will detect high energetic astrophysical phenomena that may be the product of interaction of the black hole-torus system.
The possible applications of our results can be used  in the phenomena which  are offered by a source in  microquasars and GRBs.
The perturbations of these systems could be  consequence of a matter coming from the outer boundary like reported for 
$Sgr$ $A^*$.  The component of the gas cloud, falling toward to  $Sgr$ $A^*$, has been 
reported by \citet{BSAGGFEE}. It is seen from this report that the gas cloud is captured and will accreate on $Sgr$ $A^*$ and perturb 
the possible accretion disk, which could be a shock cone \citep{DZR} or torus \citep{MosDasCz},
 around the black hole in our galactic center.  Such a perturbation may explain the properties of the flare emission
observed in the innermost part of accretion disk in the center of our galaxy.
In addition, the perturbation which has a localized density and is similar to  those of a  dense stellar wind 
may help us to understand the internal structure of  super-giant fast X-ray transient outbursts.

%%%%%%%%%%%%%%%%%%%%%%%%%%%%%%%%%%%%%%%%%%%%%%%%%%%%%%%%%%%%%%%%%%%%%%%
%%%%%%%%%%%%%%%%%%%%%%%%%%%%%%%%%%%%%%%%%%%%%%%%%%%%%%%%%%%%%%%%%%%%%%%
%%%%%%%%%%%%%%%%%%%%%%%%%%%%%%%%%%%%%%%%%%%%%%%%%%%%%%%%%%%%%%%%%%%%%%%
%%%%%%%%%%%%%%%%%%%%%%%%%%%%%%%%%%%%%%%%%%%%%%%%%%%%%%%%%%%%%%%%%%%%%%%
\section{Numerical Results}
\label{Numerical Results}

%%%%%%%%%%%%%%%%%%%%%%%%%%%%%%%%%%%%%%%%%%%%%%%%%%%%%%%%%%%%%%%%%%%%%%%

The numerical results given in this paper are classified depending on the torus-to-black hole 
mass ratio and black hole spin  as well as the  size of the pressure-supported stable torus around the 
black hole.  
The perturbation will cause to  the oscillation and dynamical change of the torus 
during the evolution.

%\noindent
We adopt the perturbation from the outer boundary at $3.1 < \phi <3.18 $
between $t=0M$ and $t=80M$ for all models except model $P_2$, which has
a continues injection during the evolution,  with different parameters
given in Table \ref{table:Initial Models1} and
the rest-mass density of perturbation is  $\rho_p =0.9\rho_c $ for all models. 
The perturbation-to-torus mass ratio is $M_p/M_{t} \sim 0.00001$. 
The matter is injected toward 
the torus-black hole system with radial velocities, $V^r = 0.002$, $V^r = 0.02$, 
$V^r = 0.2$, and  $V^r = 0.0001$ and  
with an angular velocity, $V^{\phi} = 0.00001$ for all $V^r$, 
accelerated radially and reached the torus around $t=3000M$. 
The ratio of  angular velocity of the perturbed matter to Keplerian angular velocity is 
$V^{\phi}/\Omega_K = 1/35$ at the 
outer boundary, $r=200M$. We use such a small angular velocity for perturbation due to the technical 
difficulties. If the perturbing stream carries 
significant angular momentum, 
the code crashes due to low values of atmosphere's parameters.
The perturbation which is rotating with a high angular velocity and has a very strong shock  
could crush the code.
While the period of matter at the center of torus is around 
$T_c = 151M$ for models $P_3$, $P_4$, $P_6$ and $P_8$, it is $T_c = 785M$ for models $P_1$ and $P_2$. 
The perturbation starts to influence the torus after  $\sim 16$ orbital period of 
the torus which has highest density around $r \approx 9M$ for models  $P_3$, $P_4$, $P_6$ and $P_8$. 
But it is $\sim 4$ orbital period for models
$P_1$ and $P_2$ given in in Table \ref{table:Initial Models1}. These orbital periods are measured at the location 
 where the rest-mass density  is  maximum in the unperturbed torus.

%%%%%%%%%%%%%%%%%%%%%%%%%%%%%%%%%%%%%%%%%%%%%%%%%%%%%%%%%%%%%%%%%%%%%%%
\subsection{Perturbed Torus Around the Non-Rotating Black Hole}
\label{Perturbed Torus}

%%%%%%%%%%%%%%%%%%%%%%%%%%%%%%%%%%%%%%%%%%%%%%%%%%%%%%%%%%%%%%%%%%%%%%%

%The perturbations sent into the black hole-torus system trigger the matter falling into
%the black hole due to the instability causing the sudden X-ray flare. The observed 
%$X-rays$ which hit the local gas of the torus produce light echo.  Because the torus around the 
%non-rotating and rotating black holes may shrink and  come closer to the black holes. 
%Hence, it constitutes more hot region to generate higher X-ray fluxes. 

\begin{figure*}
 \center
\psfig{file=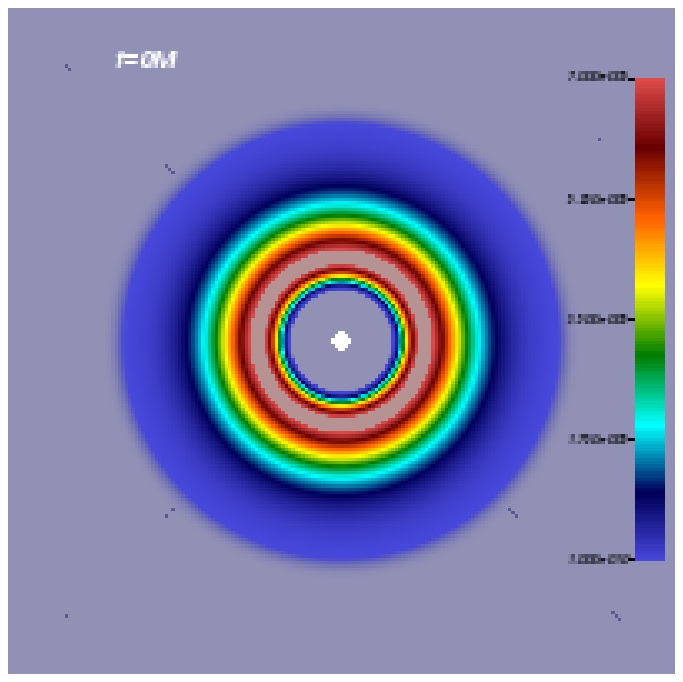,width=4.0cm}
\psfig{file=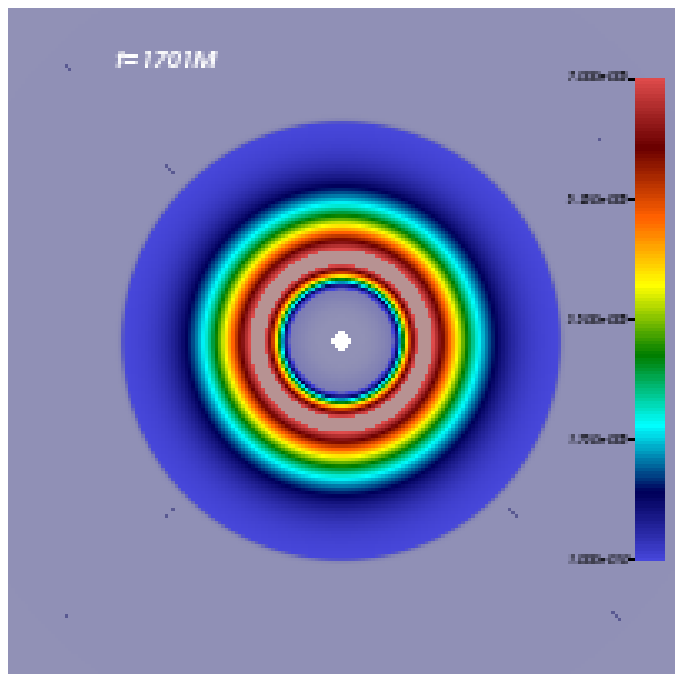,width=4.0cm}
\psfig{file=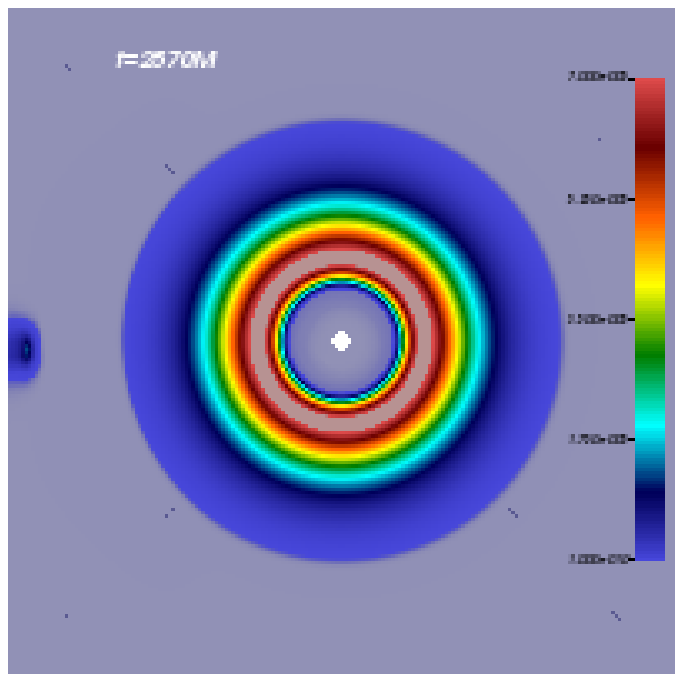,width=4.0cm}
\psfig{file=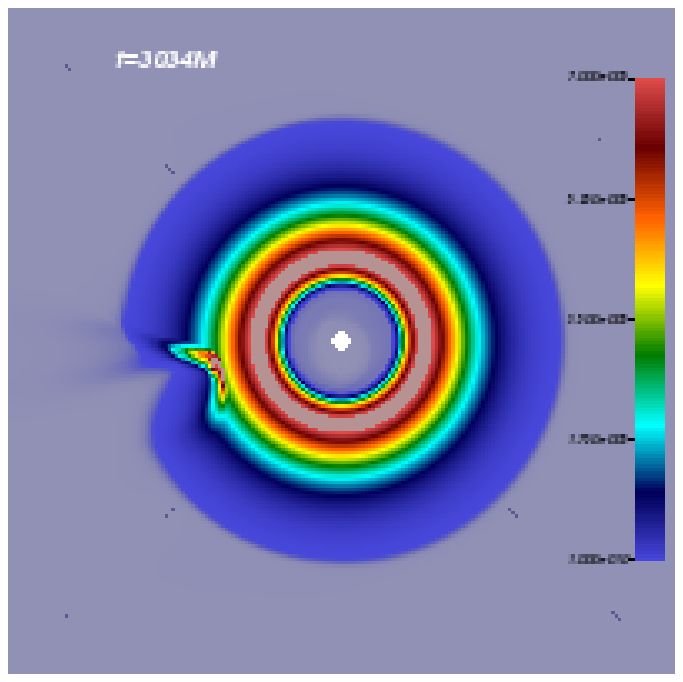,width=4.0cm}
\psfig{file=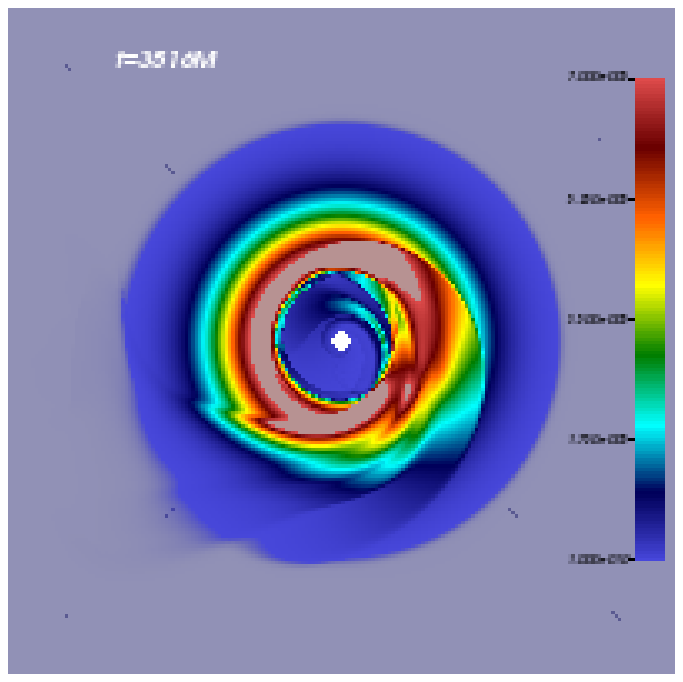,width=4.0cm}
\psfig{file=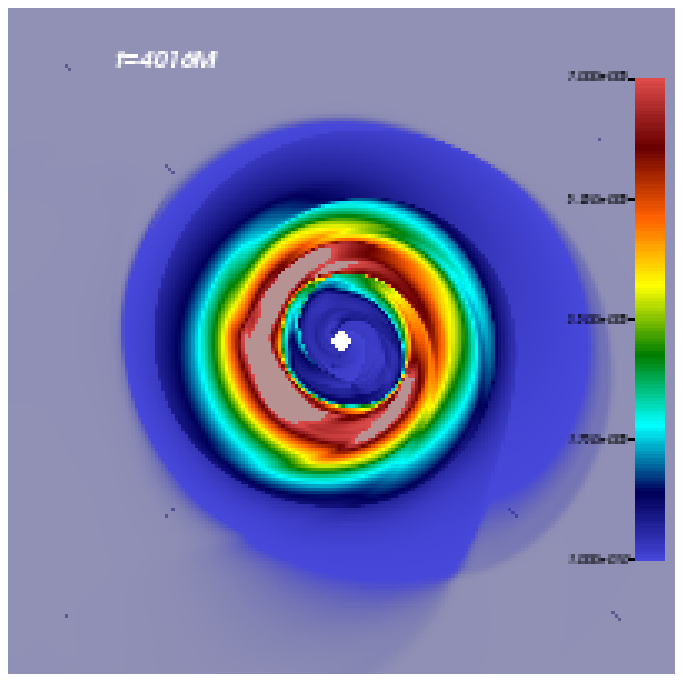,width=4.0cm}
\psfig{file=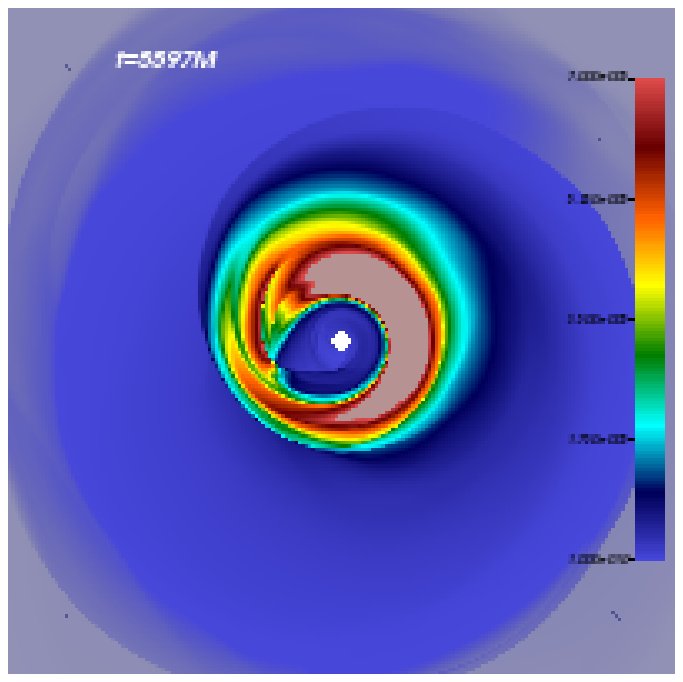,width=4.0cm}
\psfig{file=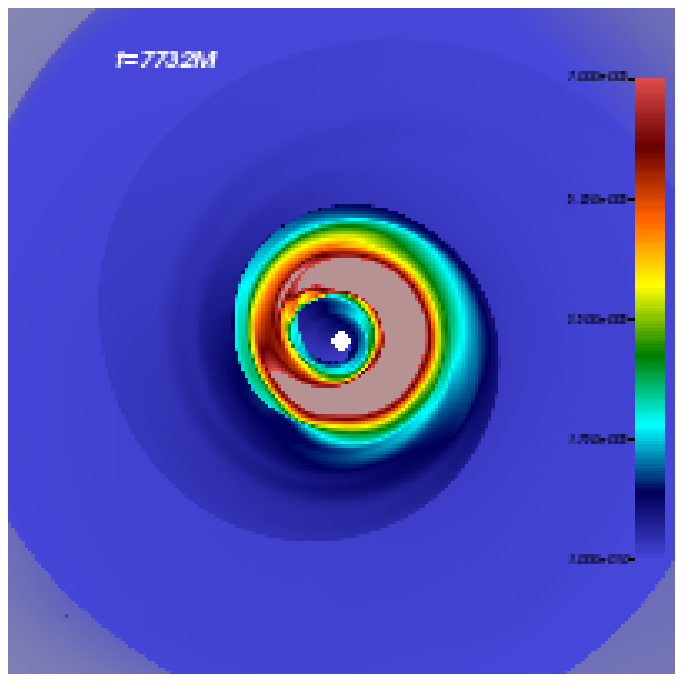,width=4.0cm}
\psfig{file=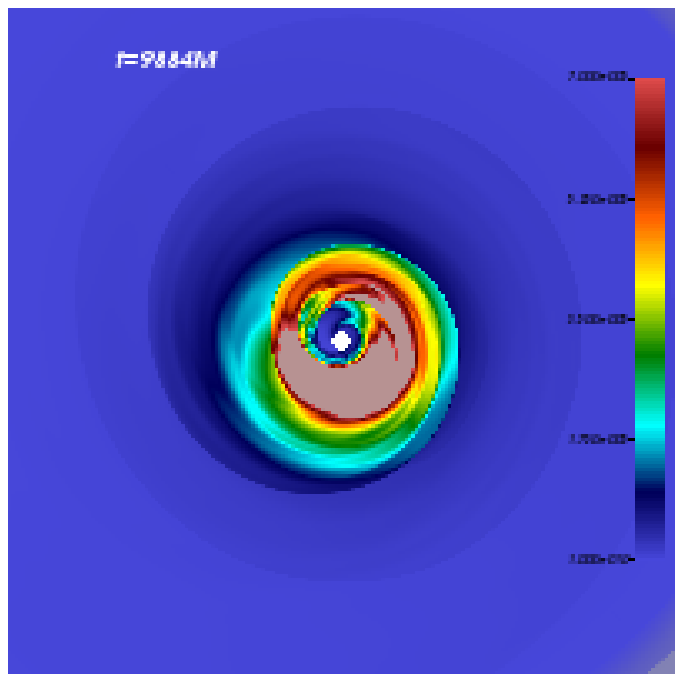,width=4.0cm}
\psfig{file=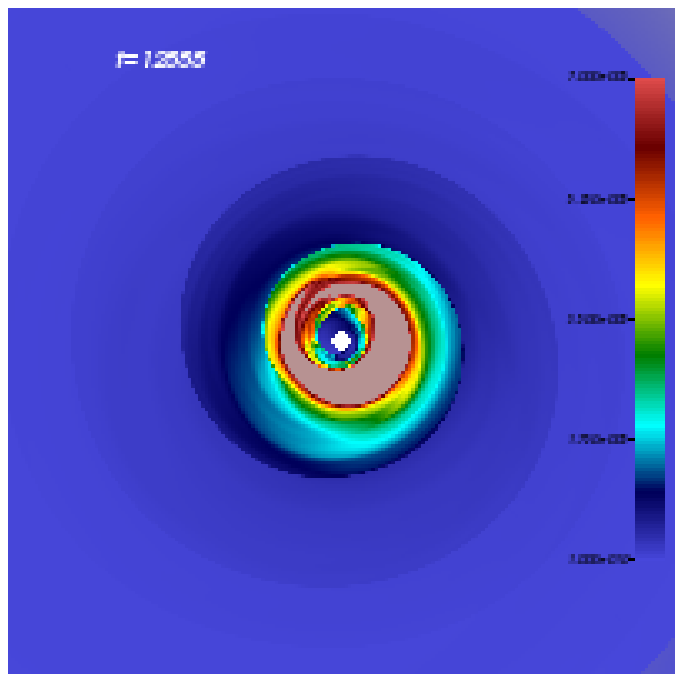,width=4.0cm}
\psfig{file=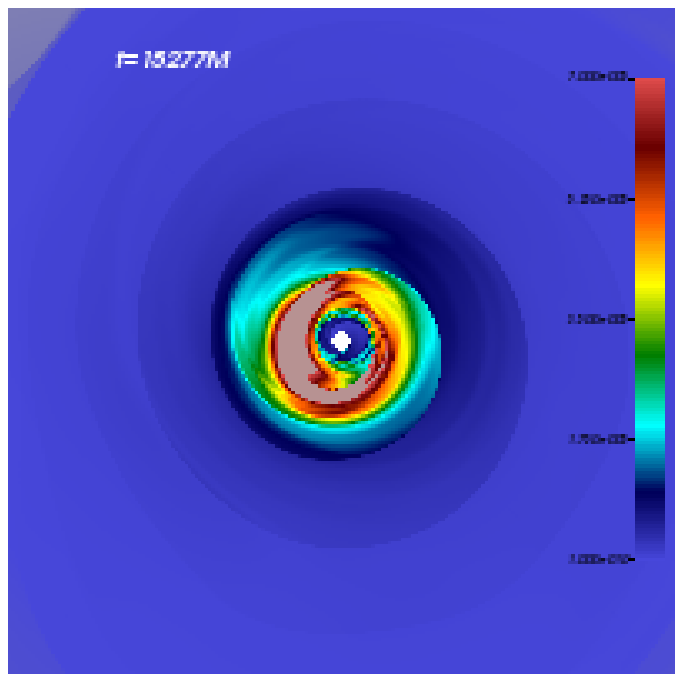,width=4.0cm}
\psfig{file=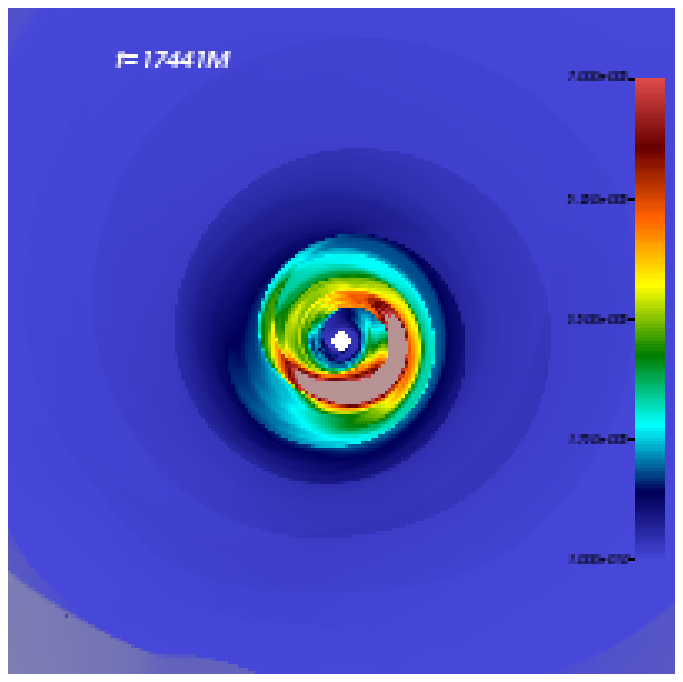,width=4.0cm}
\psfig{file=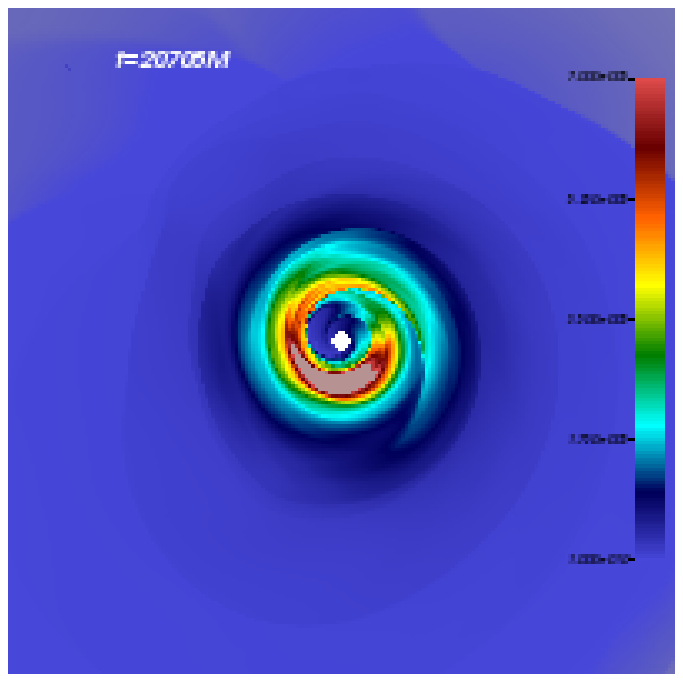,width=4.0cm}
\psfig{file=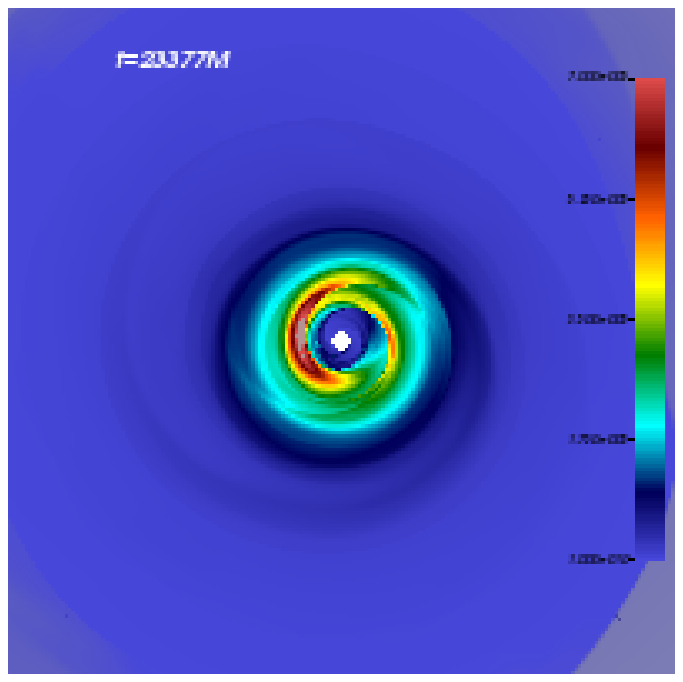,width=4.0cm}
\psfig{file=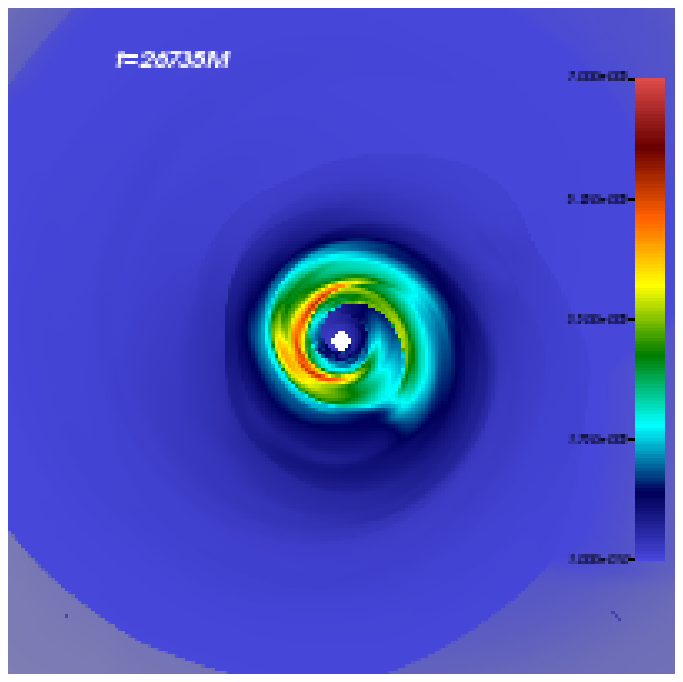,width=4.0cm}
\psfig{file=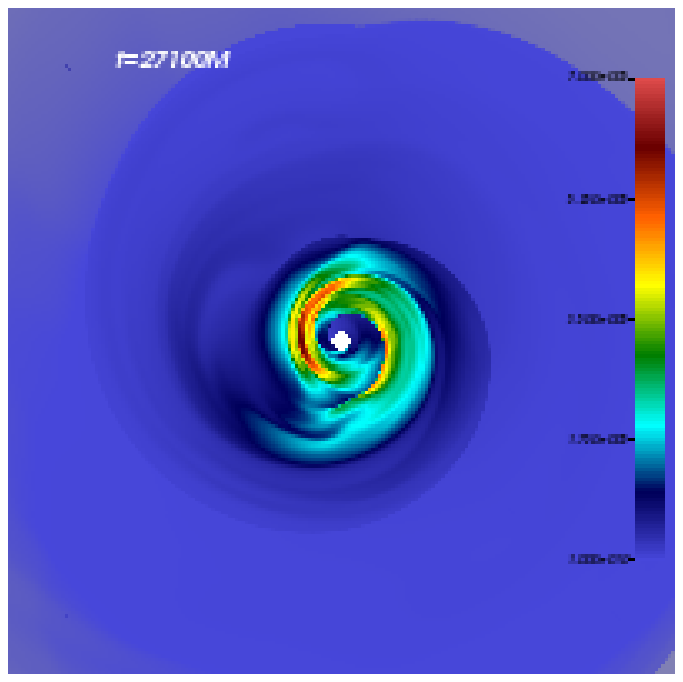,width=4.0cm}
\caption{The rest-mass density of the perturbed torus during the 
time evolution for Model $P_1$, 
given in a linear color scale.
The torus is perturbed by a matter 
coming from the outer boundary. The matter interacts with the torus 
around $t=3034M$ and distorts it. The local Papaloizou-Pringle instability is formed around
$t=4540M$ and it is suppressed during the  evolution. The structure of 
the torus is highly turbulent and makes quasi-periodic oscillation.
The domain is $[X_{min},Y_{min}] \rightarrow [X_{max},Y_{max})] =[-120M,-120M)]\rightarrow [120M,120M]$.}
%\vspace{1cm}
\label{Model P1_2}
\end{figure*}

\begin{figure}
 \center
\vspace{0.3cm}
\psfig{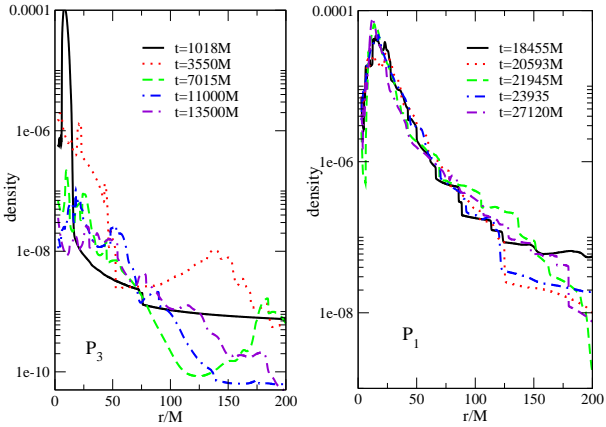}
\caption{Time evolution of the rest-mass density along the r for model 
$P_3$ (left panel) and model $P_1$ (right panel). 
The portion of the density around the cusp location is also given for 
model $P_1$.
\vspace{0.3cm}
\label{density_P1_P5}}
\end{figure}

\begin{figure}
 \center
\vspace{0.3cm}
\psfig{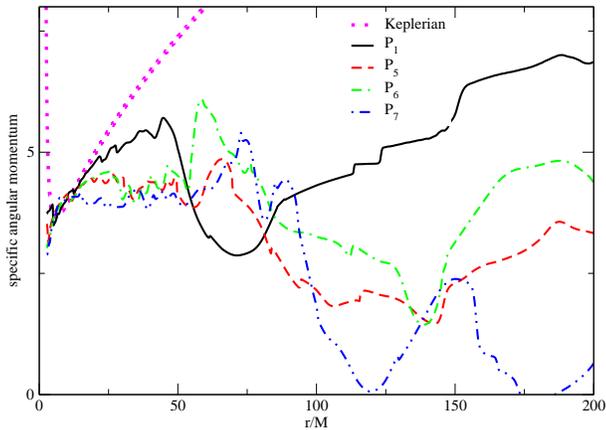}
\caption{The specific angular momentum for models $P_1$, 
$P_5$, $P_6$, and $P_7$ and for Keplerian  are plotted as a function of radial coordinate 
a long time later that the disc is perturbed and instability is developed.
\vspace{0.3cm}
\label{Specific_angular_mom}}
\end{figure}

To analyze the dynamics of the torus and its instability after the perturbation, we plot the 
density of the torus for the inner region  at different snapshots for  
model $P_1$ seen in Fig.\ref{Model P1_2}. 
The torus, initially stable, 
is perturbed by matter coming from the outer 
boundary. In the color scheme while the red is representing the highest density,
the blue shows the lowest value of the density of torus. At $t \sim 3000M$, perturbation 
hits the pressure-supported stable torus and then the spiral structure with a quasi-steady state case 
is developed as time progresses. Once the perturbed torus reaches the quasi-steady state, the 
non-axisymmetric  dynamical features are introduced in the accreted torus.
This quasi-steady state structure rotates around the black hole and 
produces quasi-periodic oscillation. 
The rotating gas represents wave-like behavior and it travels inward or outward.
From Fig.\ref{Model P1_2},  we conclude that the torus stays in steady state and the inner radius
is located around $r=15M$ before it interacts with perturbation. 
After the perturbation, 
the inner radius of the torus gets closer to the black hole and, at the same time, 
the cusp located in the equipotential surfaces of the effective
potential moves outwards into the torus.
As a result, the inner radius and cusp location of the newly developed quasi-steady-state 
torus equal to each other and  oscillate among the points, 
$r_{cusp}=r_{in}=4.71M$,  $3.48M$, $6.2M$, $3.63M$ and $4M$
which can be seen in the zoomed  part of Fig.\ref{density_P1_P5}, 
during the evolution.
The specific angular momentum corresponding to these cusp locations are  
$\ell=3.77M$,  $4.36M$,
$3.67M$, $4.23M$ and $4M$, respectively.
The location of the cusp of the  perturbed torus moves out from the black hole  and hence, 
the accreted torus reaches to a new quasi-equilibrium state.
In addition to evidence supplied by Fig. \ref{Model P1_2} and the right panel of Fig.\ref{density_P1_P5},
Fig.\ref{Specific_angular_mom} also shows that the angular momentum distribution of the torus 
increases outwards from $r\sim 5.25M$, which represents the location of the cusp at a fixed time, 
to the larger radii.
After all, the distribution of density 
is non-axisymmetric. 
It is also seen in the color plots that the 
perturbed torus has an oscillatory 
behavior, and the oscillation amplitude stays almost constant
during the compression and expansion of the torus.

The rotating torus has centrifugal forces which pull out gas outwardly 
around the black 
and causes less the gas accreted than the Bondi rate during the perturbation. 
The angular velocity of the torus is larger than Keplerian one, 
and it exponentially decays for the larger r. The distribution of angular momentum is maintained by 
the disc pressure. The angular momentum of the torus close to the horizon may be transferred to the black hole after 
the torus is perturbed. This transformation can be used to explain the spin up of the black hole.

In order to reveal the effect of the size of torus and inner radius onto the 
instability and dynamics of the torus, 
we consider the perturbations of the tori for the same mass, but with different sizes. 
The time evolution of the central rest-mass density of torus, each of them plotted along 
$r$ at $\phi=0$, are shown for models $P_1$ and $P_3$ in Fig.\ref{density_P1_P5} . For these
models, we have applied a perturbation, which is $\rho_p =0.9\rho_c $, from the outer boundary 
between $t=0M$ and $t=80M$ to evolute the influence of perturbation to the overall dynamics
of torus. The perturbation destroys the torus, triggers the  Papaloizou-Pringle instability and can cause  the location 
which has the maximum 
density to approach the horizon. Such perturbation induces the matter to fall into the black hole or 
out from computational domain. While the maximum rest-mass density of torus  slightly oscillates 
for model $P_1$, seen in the right panel of Fig.\ref{density_P1_P5}, the rest-mass density for model $P_3$ 
significantly decreases, seen 
in the left panel of the same graph, during the time evolution.

%

\begin{figure*}
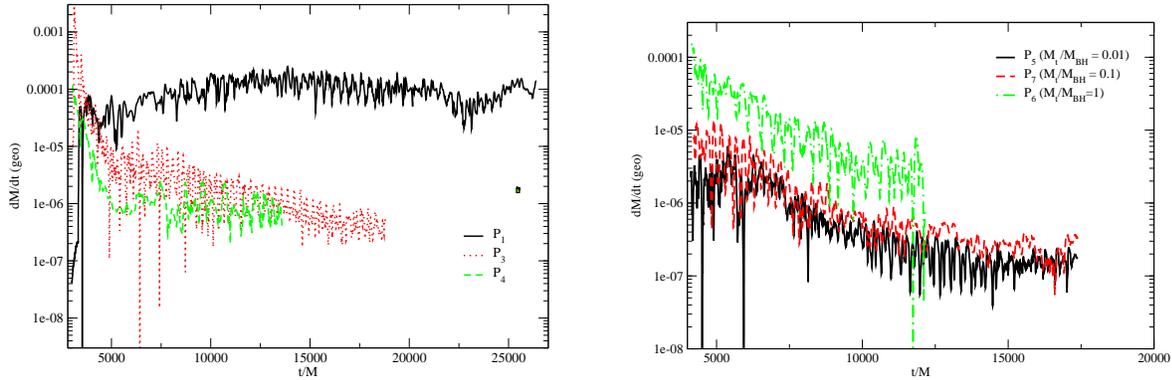

 \center
\vspace{0.5cm}
\psfig{file=P1_P3_P4_Mass_acc.eps,width=7.2cm}
\hskip 1.0cm
\psfig{file=Mass_acc_P5_P6_P7.eps,width=7.2cm}
\caption{Left panel: the mass accretion rates for models $P_1$, $P_3$, and $P_4$ computed
at $r=6M$ in geometrized unit.
Three different models, the torus with different $r_{in}-r_c$  while keeping the
 same sound velocity or vice verse,   are considered to show the effect of 
perturbation on the torus. Right panel: mass accretion rate for models  $P_5$, 
$P_6$, and $P_7$.
\label{Mass_acc1}}
\end{figure*}

%\noindent
Accretion rate shows time variation in 
a great amplitude similar to one happening in low or hard X-ray time variation. 
Computing the accreting and the mass accretion rate are important indicators to understand the disc 
behavior and its instability around the black holes. 
The non-axisymmetric perturbation onto the steady-state torus, which has a constant 
specific angular momentum, around the black hole can describe the Papaloizou-Pringle instability.
Instability of the accreted torus can be subject to 
perturbation coming from the outer boundary in a number of circumstances which may 
cause unstable mass accretion rates \citep{PapPri1, PapPri2}. 
We have confirmed that the accretion rate 
is governed by some physical parameters: the torus-to-hole mass ratio, the location of 
cusp, initial specific angular momentum of the torus, Mach number of the perturbed matter
and the angular momentum of the black hole. The mass accretion rate computed
from $P_1$, $P_3$, and $P_4$ are depicted in the left part of Fig.\ref{Mass_acc1}. We have noticed 
that the mass accretion rate for the models are the same at early times of simulations, but later
the mass accretion rate for models, $P_3$ and $P_4$, drop exponentially as a function of time and 
the  Papaloizou-Pringle instability is developed. 
Interestingly, the mass 
accretion rate for only model $P_1$  does not increase in amplitude, and it oscillates around 
$dM/dt = 5$ x $10^{-5}$.

The mass accretion rate
of the perturbed torus produces 
luminosity associated with the central engine of gamma ray burst. The expected maximum 
accretion rate from the gamma ray burst is as high as 
$dM/dt \approx 0.01 - 1M_{\odot}/s$ \citep{ChBe}. Fig.\ref{Mass_acc1} indicates that 
the computed mass accretion rate from our numerical simulations,  
$\frac{dM}{dt} \sim$ 2 x $10^5 (\frac{M_{\odot}}{s})\frac{dM}{dt}(geo)$
where $\frac{dM}{dt}(geo)$ is the mass accretion rate in geometrized unit, is  
in order of or slightly higher than the expected maximum accretion rate. 
It is also shown in the left panel of Fig.\ref{Mass_acc1} that 
the mass accretion rate suddenly increases after perturbation reaches to the torus and shortly
after the Papaloizou-Pringle instability is developed.
All the models in Fig.\ref{Mass_acc1} exhibit the  Papaloizou-Pringle instability which
grows exponentially with time and present a non-linear growth rate. After long time later
(it is approximately $t=15000M$ for  models $P_5$ and $P_7$), it reaches a saturation point
and finally reaches a new quasi-equilibrium point. Before the saturation point, the torus looses mass
and then it relaxes to  the quasi-stationary accretion state.

%\noindent
It is obvious from the right part of Fig.\ref{Mass_acc1} that the 
accretion rates exponentially decay during the time and is always 
bigger when the mass ratio of the torus-to-black hole is larger. Decaying a mass
accretion rate reaches a constant value $dM/dt \approx 0.1 M_{\odot}/s$ 
around $t=14000M ( t=82 t_{orb})$. It is in good agreement with the expected 
accretion rates for accreting disc with the GRBs.
As it can be seen in Fig.\ref{Mass_acc1},   the amplitude of 
mass accretion rate and its behavior manifest a dependence on 
the mass ratio of the torus-to-black hole.  While 
the mass accretion rates for  all models decay with time due to the strong gravity 
which  overcomes angular momentum of the torus,  it oscillates around 
a certain rate for model $P_1$, shown in Fig.\ref{Mass_acc1}. The oscillation
behaviors of all models are seen  during the time evolution.  
When the accreted torus gets close to the black hole and cusp
moves outwards into the torus, model $P_1$ reaches persistent phase of quasi-periodic oscillation
without the appearance of  the Papaloizou-Pringle instability.

\begin{figure}
 \center
\vspace{0.3cm}
\psfig{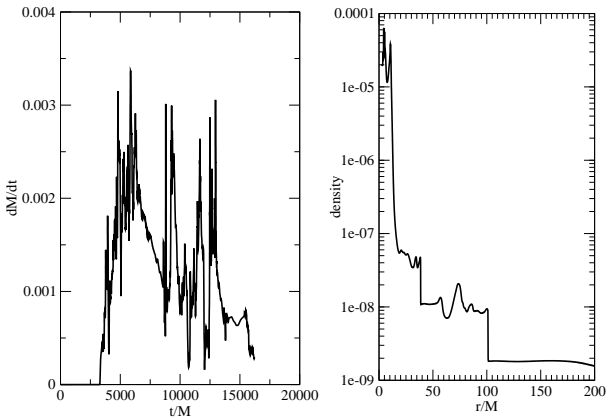}
\caption{The mass accretion rate (left panel) and 
the rest-mass density of torus along r (right panel) are plotted 
for model  $P_2$ .
\vspace{0.3cm}
\label{variables_P2}}
\end{figure}

%\noindent
The continuous  perturbation of the black hole-torus system can lead not only instability of the torus
but also oscillates of its dynamics around the certain points, seen in Fig.\ref{variables_P2}.
The left panel in Fig.\ref{variables_P2} exhibits the time variation of mass accretion rate
for model $P_4$ computed for $r=8M$ which is true for the perturbed matter
injecting from the outer boundary continuously. 
The density of the torus at a fixed time $t=17200M$ and $\phi=0$ along $r$  is shown in the right panel.
It is seen in this plot that the perturbation starts to 
interact with a torus around $t \sim 3000M$, and then nonlinear oscillation is developed.  
It is noted 
that the continuous perturbation can create different types of dynamical structures, accretion
rates and shock wave on the disk.


Accretion into the black hole or out from the computational domain  is driven 
by a transport of angular momentum via spiral shock waves created on the torus.
The higher specific angular momentum of the torus outwards can allow
the torus to become more stable. Fig.\ref{Specific_angular_mom} shows the specific 
angular momentum of the torus along the radial coordinate for models $P_1$, 
$P_5$, $P_6$, and $P_7$. The Keplerian specific angular momentum is also plotted at the top 
of these models to compare them. The specific angular momentum changes drastically depending on the  radius.
It becomes apparent that the increasing of the specific angular 
momentum outwards leads to stabilize the torus and suppresses the
Papaloizou-Pringle instability, seen in Fig.\ref{Model P1_2}. The Papaloizou-Pringle instability visible  
immediately after the 
perturbation starts to influence of the stable torus. 
The Papaloizou-Pringle instability performs a torque on the torus and the angular momentum of the torus
is redistributed. So it amplifies the mass accretion rate \citep{ZurBen}.
The rotating quasi-steady-state torus can only be seen in model  $P_1$. Because the 
substantial amount of the specific angular momentum is kept inside the region $r<50M$.
Meanwhile, it is seen in Fig.\ref{Specific_angular_mom} that specific angular momentum is 
sub-Keplerian which indicates that the gas pressure creates a pressure force and it supplies a radial 
support to the torus.

The newly developed angular momentum distribution of the black hole-torus system, after the 
non-axisymmetric perturbation, might have the local Rayleigh stability condition depending on the 
slopes and its signs.  As you can see in 
Fig.\ref{Specific_angular_mom},  the slope of the specific angular momentum is bigger 
than zero ($d\ell/dr>0$), which is the limit of  the Rayleigh-stable torus, 
in model $P_1$ for $r <50M$, but it is  $d\ell/dr<0$ and produces  the Rayleigh-unstable torus 
for $50M<r<75M$. Any increase of the specific angular momentum in  the radial direction 
stabilizes the torus.

\begin{figure}
 \center
\vspace{0.3cm}
\psfig{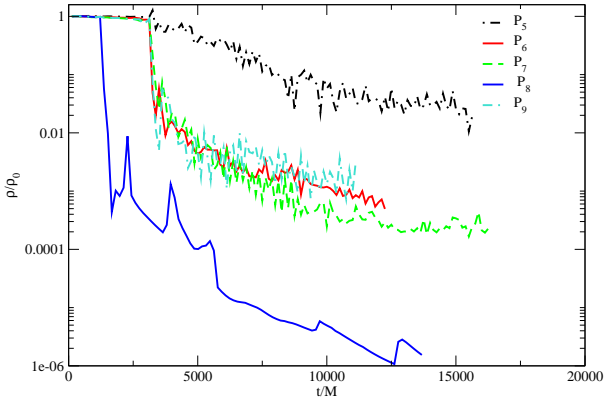}
\caption{ The variation of the maximum density of torus  in logarithmic scale
as a function of time, each of them normalized to its initial value 
$\rho_o \equiv \rho_c$, 
for the models, $P_5$, $P_6$, $P_7$,  $P_8$, and $P_9$. The torus loses the matter during the 
time evolution for all cases while it is almost stable before the 
perturbed matter reaches the torus.
\vspace{0.3cm}
\label{max_density_P5_P6_P7}}
\end{figure}

Fig.\ref{max_density_P5_P6_P7} represents the time evolution of the maximum rest-mass 
density at the center of the torus, each of them normalized to their initial values, for 
models $P_5$, $P_6$, $P_7$, $P_8$, and $P_9$. For these models, we have applied the same perturbation 
with different densities ($\rho_p =0.9\rho_c $), giving the first kick to the 
black hole-torus system. The matter begins to influence the torus around $t=3120M$. 
The perturbation triggers the non-stable oscillation of the 
torus for models given in Fig.\ref{max_density_P5_P6_P7} except the model $P_8$.   These models 
exponentially decay and produce sharp peaks after perturbation reaches the torus.
Thus, these results imply that subsonic or mildly supersonic perturbation  produces 
the pressure-supported oscillating torus around the black hole, and amplitude of the oscillation 
gradually decreases while the Mach number of perturbation increases.  
Such oscillations can cause the matter to fall into the black hole or outward from the computational 
domain. These physical phenomena can significantly reduce the rest-mass density of the torus 
during the time evolution seen in  Fig.\ref{max_density_P5_P6_P7}. 
Depending on the initial configurations, such as, mass ratio of the torus-to-hole and 
inner radius of the last stable orbit of the torus, the reducing rate of
the rest-mass density  is defined at the end of simulation. The density of the stable initial 
torus having an initial radius 
$r_{in} = 5.49M$ given in model $P_5$  decreases slower than the one with the smaller inner 
radius given in models  $P_6$, $P_7$, $P_8$, and $P_9$. This result is also 
confirmed by model  $P_1$. The matter of the  
torus close to the black hole would fall into the black hole due to the  strong  gravity after the 
torus is triggered by perturbation. Therefore,  the mass of the black hole and 
its spin would increase
with time significantly. We also note that the reduction of the rest-mass density for model   
$P_6$  is slightly slower than the model $P_7$.  So the lower the torus-to-hole mass ratio 
looses the faster the matter during the evolution. But it needs a further consideration 
to reveal a global conclusion. 

%

%%%%%%%%%%%%%%%%%%%%%%%%%%%%%%%%%%%%%%%%%%%%%%%%%%%%%%%%%%%%%%%%%%%%%%%%%%%%%%%%%%%%%%%%%%%%%%
%%%%%%%%%%%%%%%%%%%%%%%%%%%%%%%%%%%%%%%%%%%%%%%%%%%%%%%%%%%%%%%%%%%%%%%%%%%%%%%%%%%%%%%%%%%%%
\subsection{Perturbed Torus Around the Rotating Black Hole}
\label{Perturbed Torus for rotating}
\begin{figure*}
 \center
\psfig{file=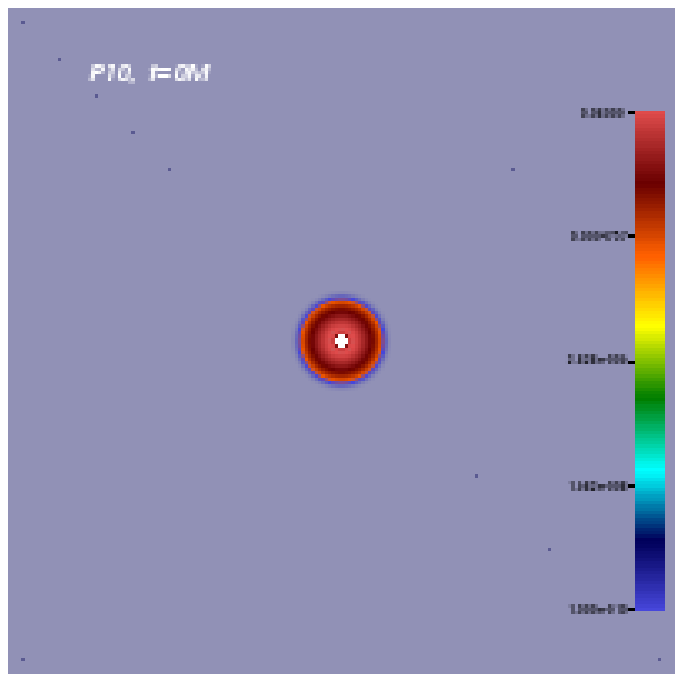,width=4.0cm}
\psfig{file=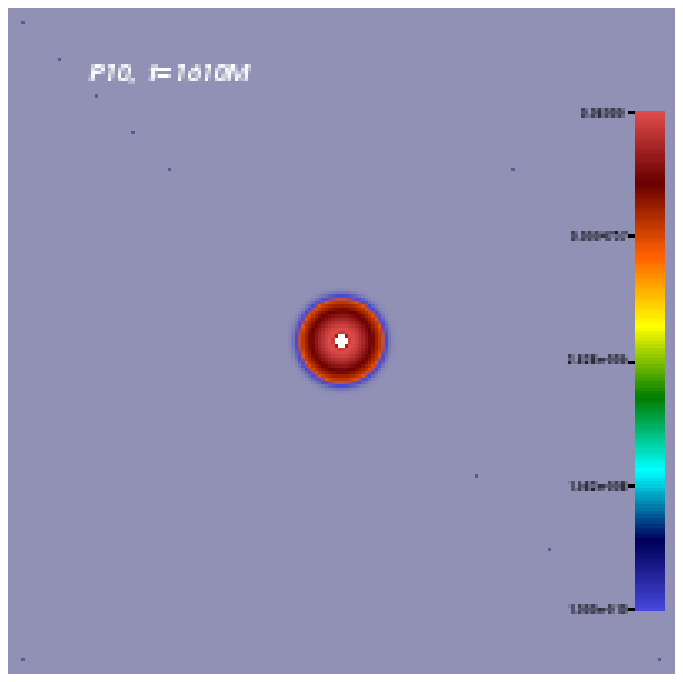,width=4.0cm}
\psfig{file=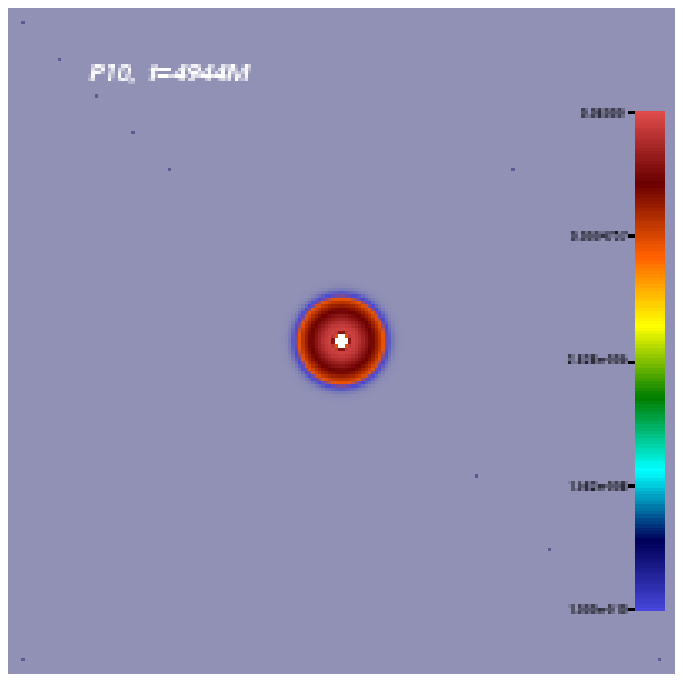,width=4.0cm}
\psfig{file=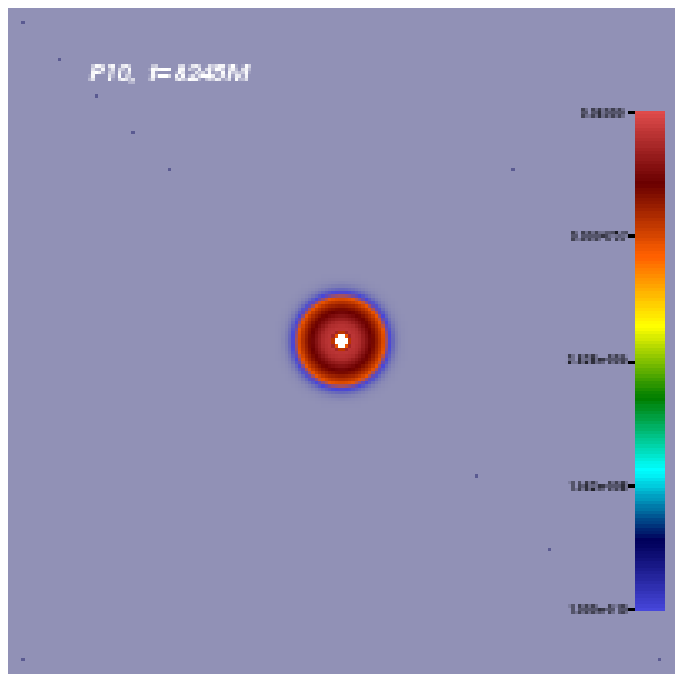,width=4.0cm}
\psfig{file=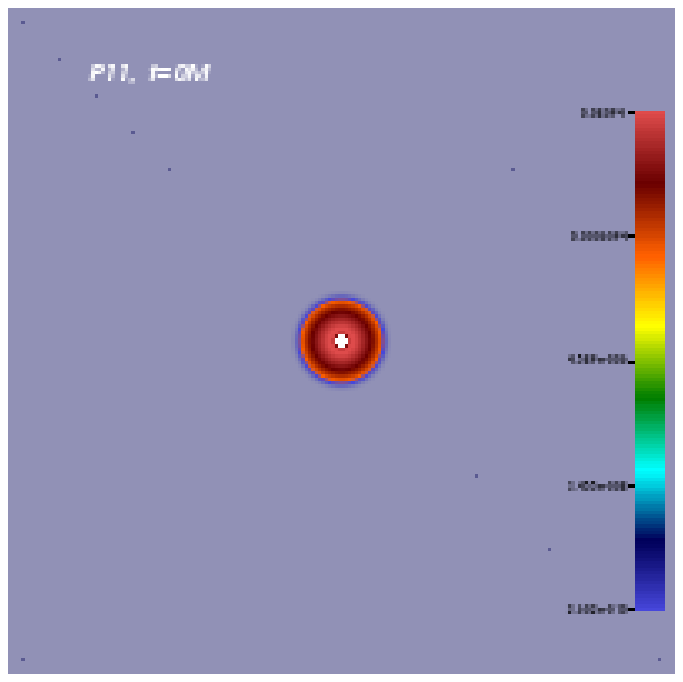,width=4.0cm}
\psfig{file=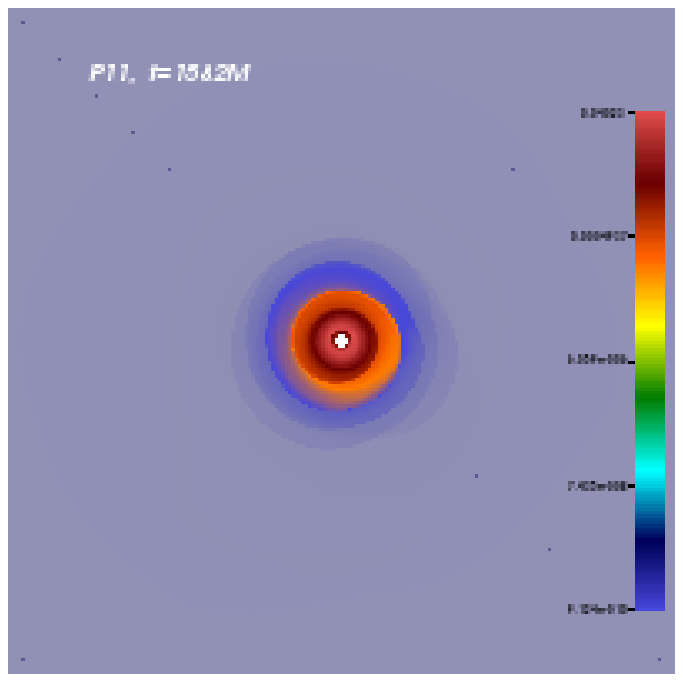,width=4.0cm}
\psfig{file=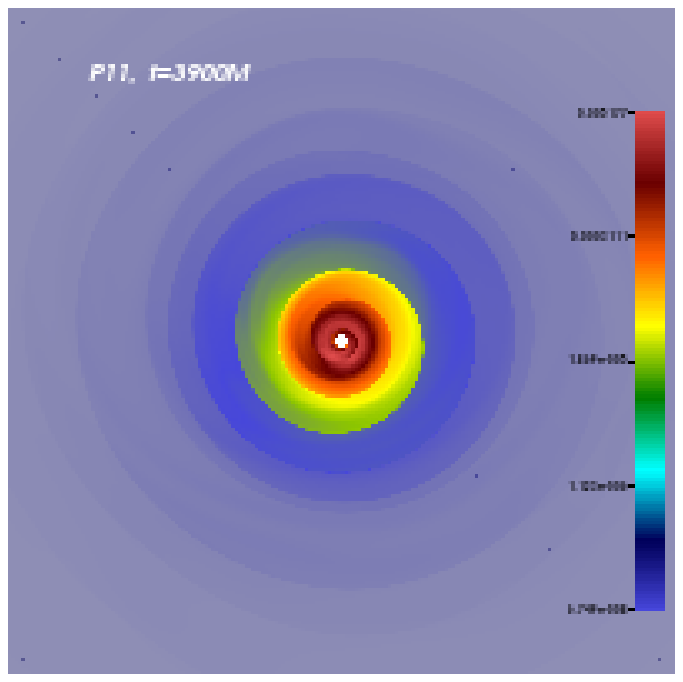,width=4.0cm}
\psfig{file=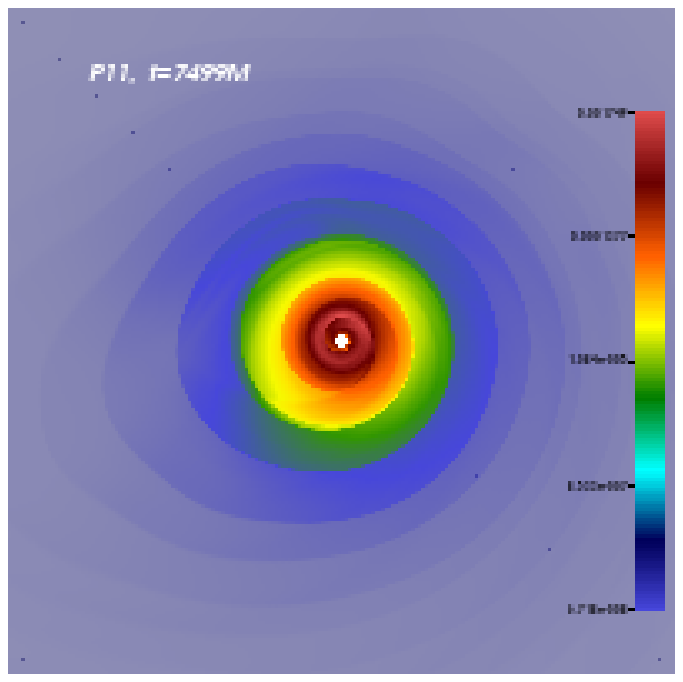,width=4.0cm}
\psfig{file=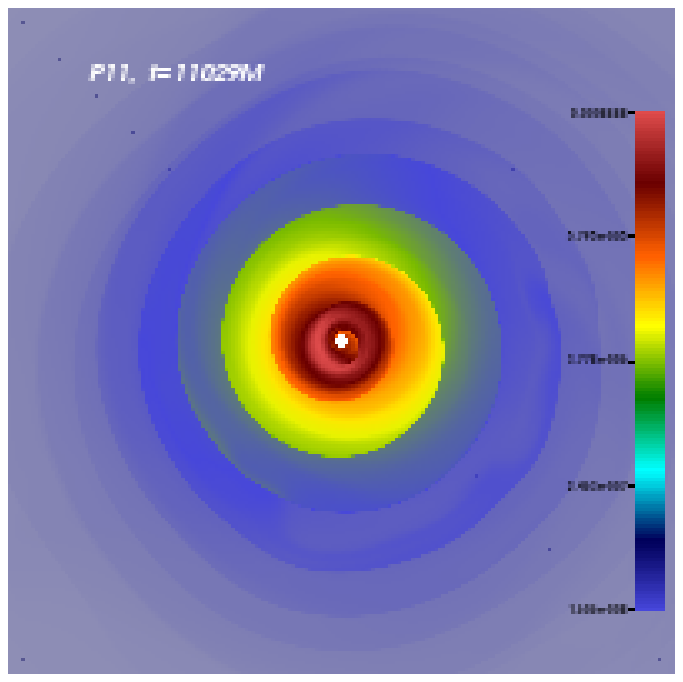,width=4.0cm}
\psfig{file=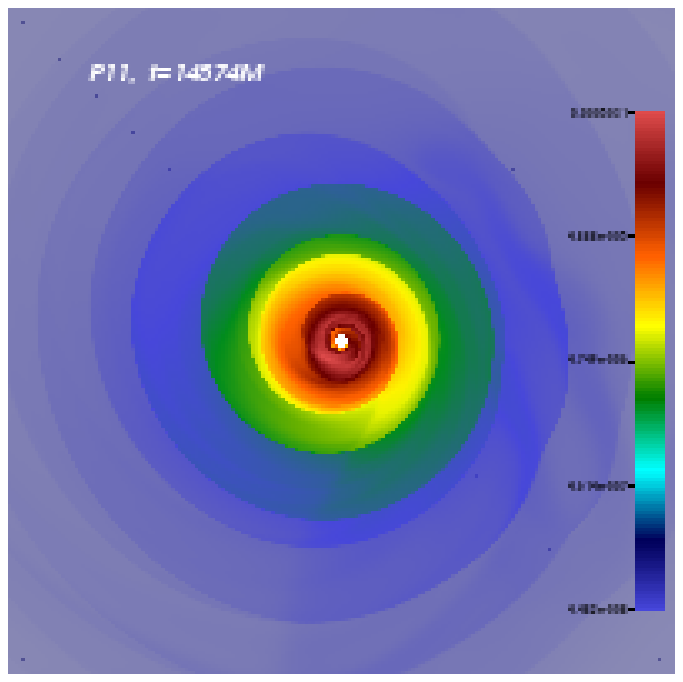,width=4.0cm}
\psfig{file=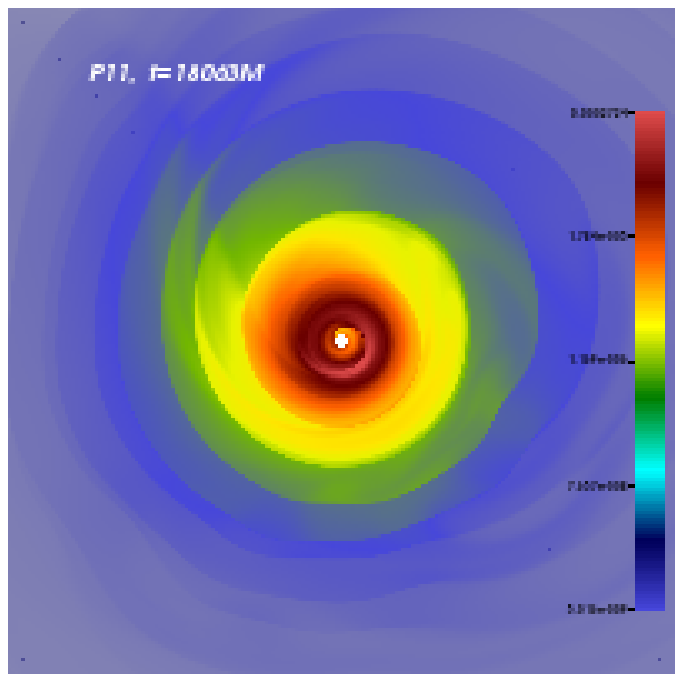,width=4.0cm}
\psfig{file=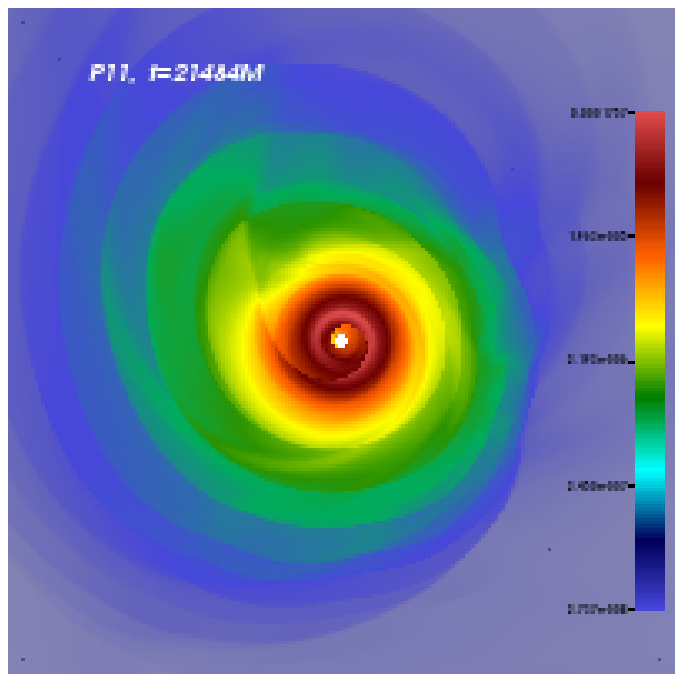,width=4.0cm}
\psfig{file=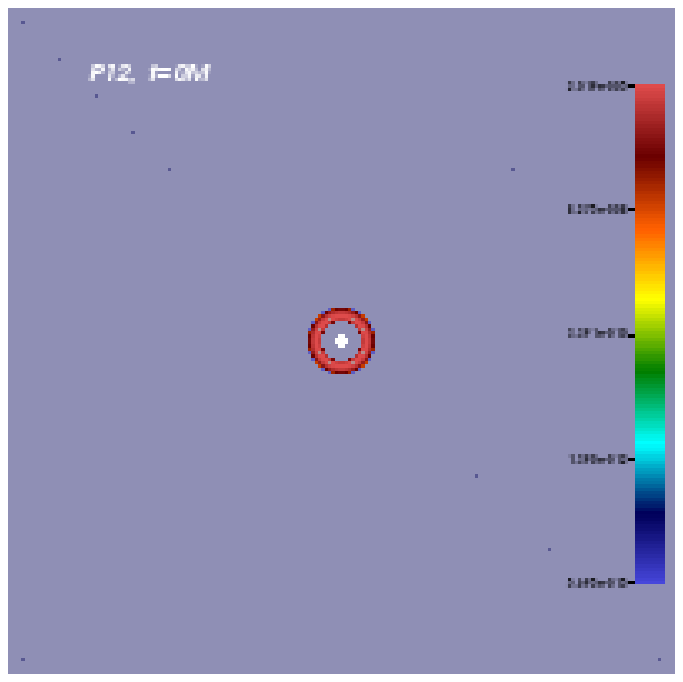,width=4.0cm}
\psfig{file=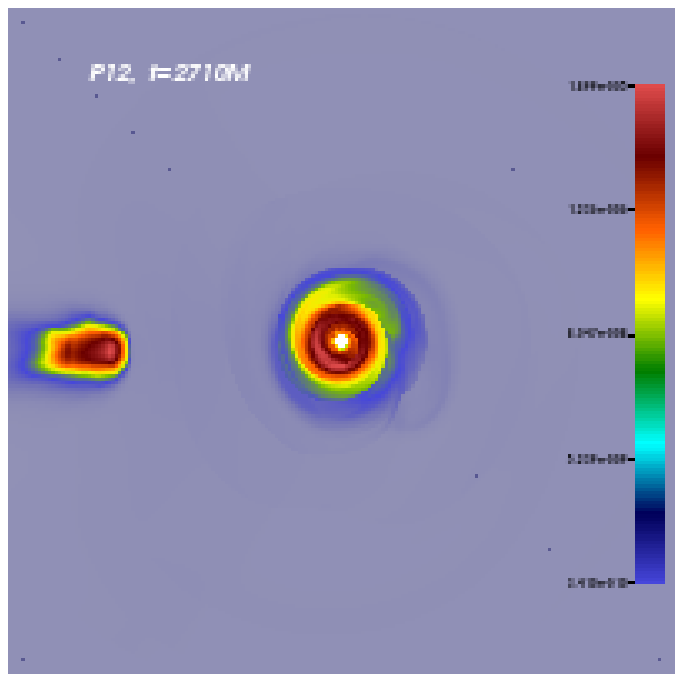,width=4.0cm}
\psfig{file=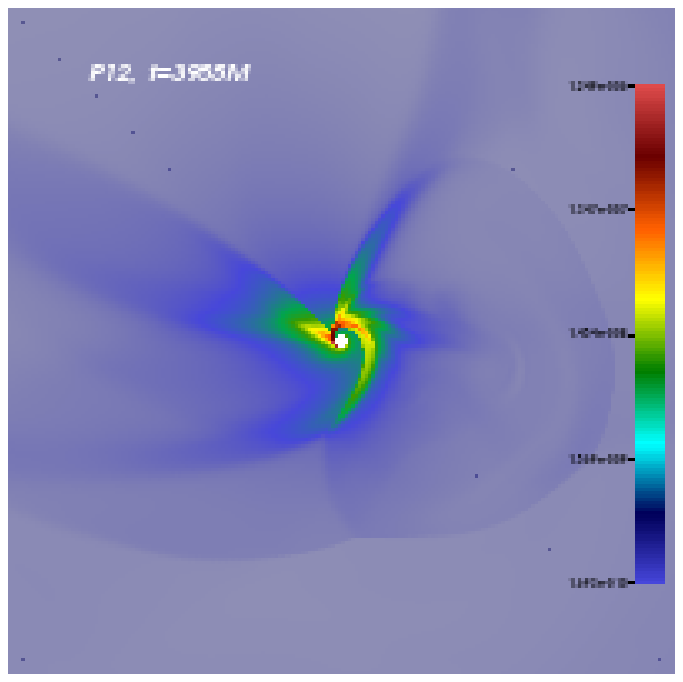,width=4.0cm}
\psfig{file=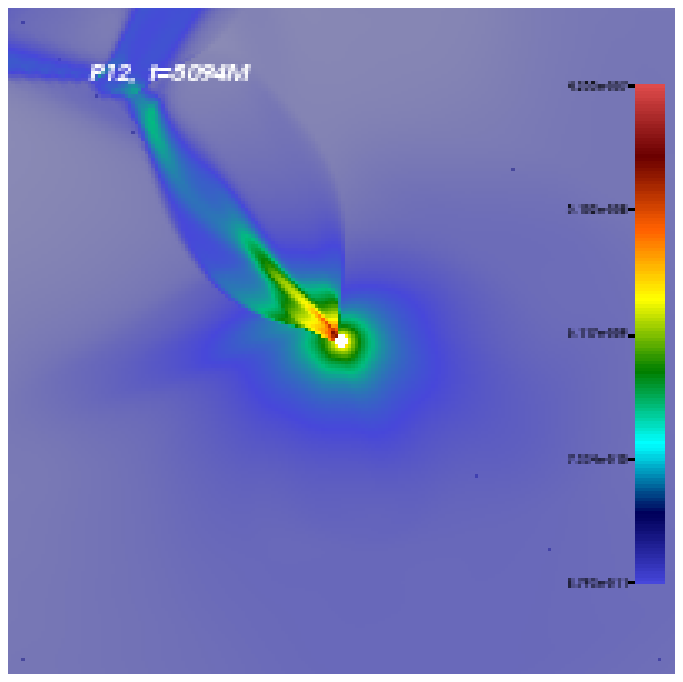,width=4.0cm}
\psfig{file=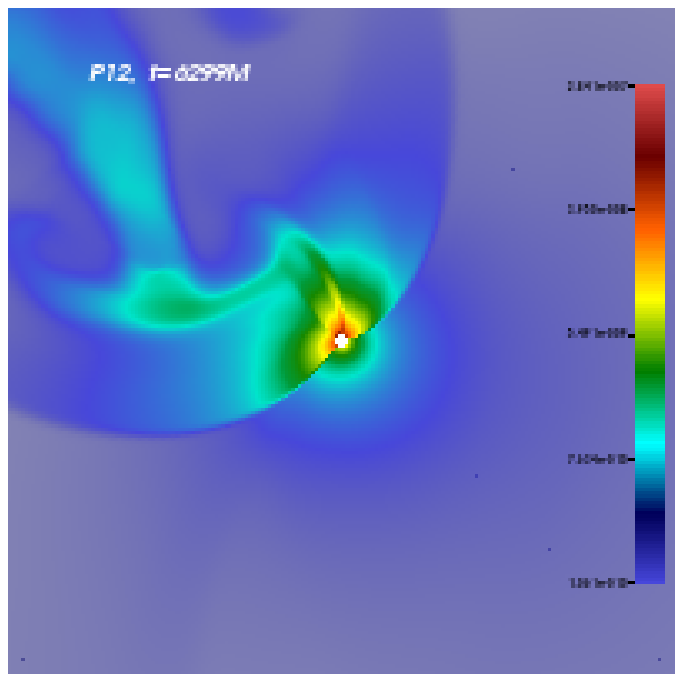,width=4.0cm}
\psfig{file=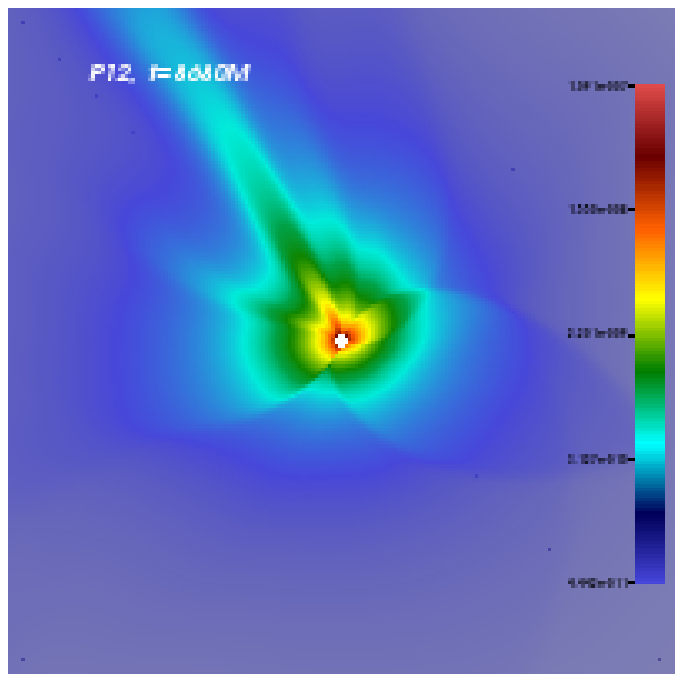,width=4.0cm}
\psfig{file=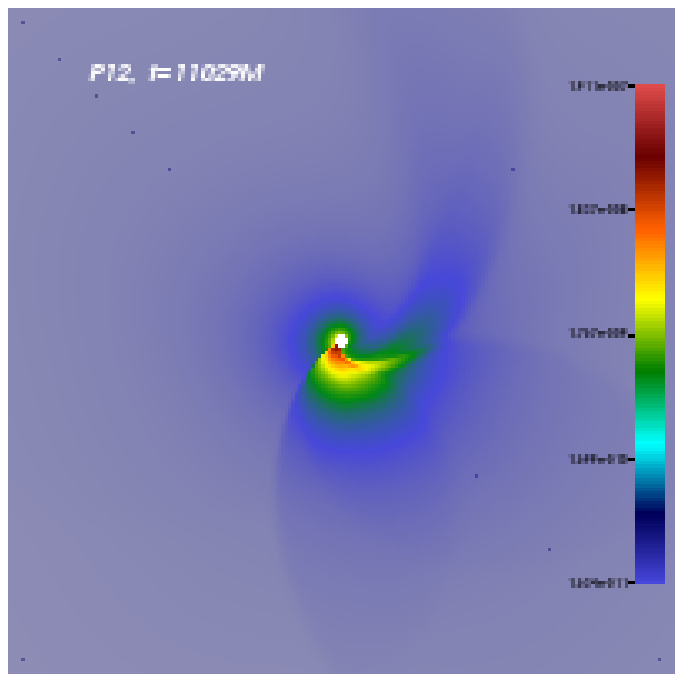,width=4.0cm}
\psfig{file=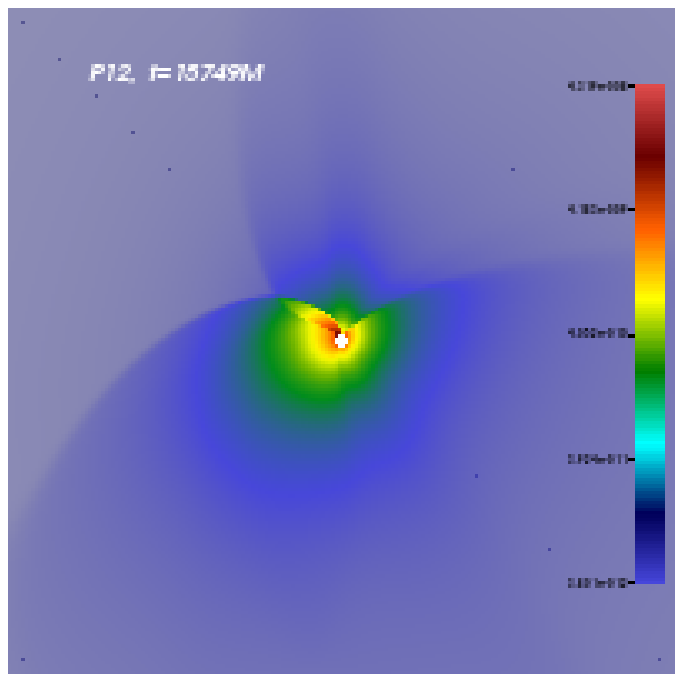,width=4.0cm}
\caption{Density $\rho$ of the torus at the equatorial plane for different models,
$P_{10}$, $P_{11}$, and $P_{12}$ with a black hole rotation parameter, $a=0.9$,
given in a linear color scale.
While the top row shows the unperturbed torus around the rotating black hole, the others 
indicate the perturbation of torus having different size and density.
The domain is $[X_{min},Y_{min}] \rightarrow [X_{max},Y_{max})] =[-100M,-100M)]\rightarrow [100M,100M]$.}
%\vspace{1cm}
\label{Model a_09}
\end{figure*}

%%%%%%%%%%%%%%%%%%%%%%%%%%%%%%%%%%%%%%%%%%%%%%%%%%%%%%%%%%%%%%%%%%%%%%%%%%%%%%%%%%%%%%%%%%%%%
In order to put forth the effect of the rotating black hole onto the perturbed torus, we have perturbed
the stable torus orbiting around  the rotating black hole using  the initial parameters given in models,  
$P_{10}$, $P_{11}$, and $P_{12}$ in Table \ref{table:Initial Models1}. We have analyzed the time 
evolution of the rest-mass density of the torus with/without perturbation 
and found that perturbed torus given 
in model $P_{12}$ become unstable before the perturbation reaches to it 
but it is stable for model  $P_{11}$. The period of fluid at the location of the highest density 
for model $P_{11}$ is $T_c=39.4M$.
The structure of unperturbed torus on a fixed space-time 
metric background, called 
a Cowling approximation,
never changes during the evolution shown in Fig.\ref{Model a_09}.
The top row of  Fig.\ref{Model a_09} clearly shows that the torus in the 
Cowling simulation does not develop a Papaloizou-Pringle instability where
steady state structure of torus remains almost constant during the evolution 
(i.e. at least $t \sim 11t_{orb}$). Because the negligible mass accretion rates 
through the cusp freeze the growth of the  Papaloizou-Pringle instability modes in 
Cowling approximation \citep{Hawley}.

\begin{figure}
 \center
\vspace{0.5cm}
\psfig{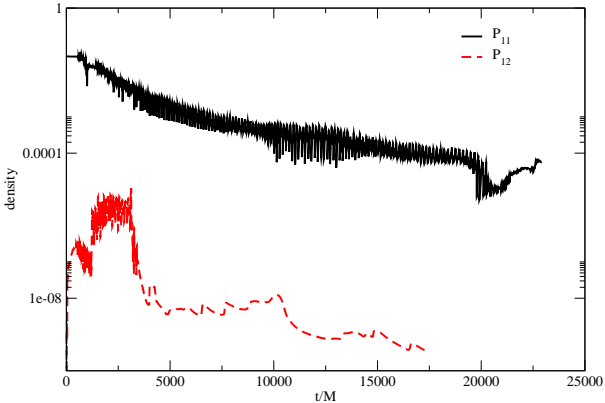}
\caption{The time evolution of the rest-mass density at a fixed point on the torus for models, 
$P_{11}$ and $P_{12}$, for $a=0.9$. The inner radii of the models are different, but the same 
perturbation is applied to the both models. 
\vspace{0.3cm}
\label{Model a_09_density}}
\end{figure}

The perturbations on the torus having different inner radii and total mass  produce diverse 
dynamics.  As it can be seen in Fig.\ref{Model a_09_density} that  the spiral pattern is produced around 
the rotating black hole, and it causes the gas falling into the black hole during the simulations. 
At the same time, it also creates regular oscillation seen in the second and third rows of Fig.\ref{Model a_09}
and  model $P_{11}$ in Fig.\ref{Model a_09_density}. It is seen from  the model  $P_{12}$ in 
Fig.\ref{Model a_09_density} and, last two rows of Fig.\ref{Model a_09} that 
the torus responds somewhat differently to the applied perturbation, and the torus with a 
low mass, which is far away from the black hole, has a non-oscillatory behavior.
Both of these models indicate that at the end of the simulation, the rest-mass density  of the torus is reduced
substantially. On the other hand, it is shown that the torus, which does not have a cusp point,
is unstable even before the perturbation, 
and has a non-linear dynamic during 
the evolution as seen in the last two rows of 
Fig.\ref{Model a_09}.

%

%%%%%%%%%%%%%%%%%%%%%%%%%%%%%%%%%%%%%%%%%%%%%%%%%%%%%%%%%%%%%%%%%%%%%%%
%%%%%%%%%%%%%%%%%%%%%%%%%%%%%%%%%%%%%%%%%%%%%%%%%%%%%%%%%%%%%%%%%%%%%%%
\subsection{Low-Density Perturbation}
\label{Low-Density Perturbation}

%%%%%%%%%%%%%%%%%%%%%%%%%%%%%%%%%%%%%%%%%%%%%%%%%%%%%%%%%%%%%%%%%%%%%%%

\begin{figure}
 \center
\vspace{0.5cm}
\psfig{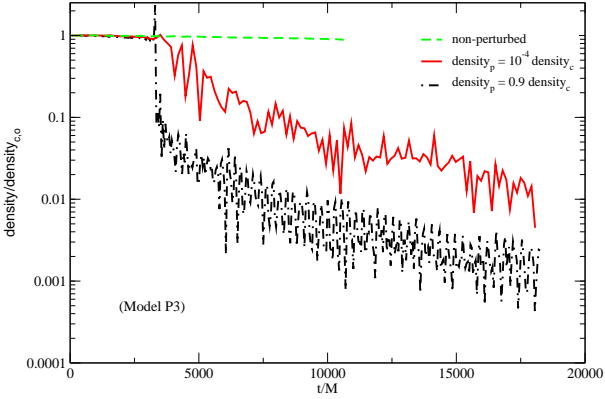}
\caption{The evolution of the rest-mass density of the torus at the point, where the rest-mass density is
maximum, for model $P_3$ with no-perturbation and with different 
values of densities of the perturbation in logarithmic scale. It is normalized to its initial value 
using the density at the computed point. 
\vspace{0.3cm}
\label{Smal_Disk_compare}}
\end{figure}

\begin{figure}
 \center
\vspace{0.3cm}
\psfig{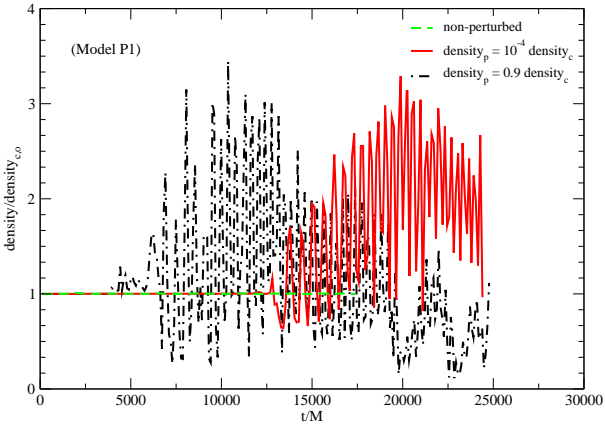}
\caption{Same as Fig.\ref{Smal_Disk_compare} but it is for model $P_1$  and linear scale is used.
\vspace{0.3cm}
\label{Big_Disk_compare}}
\end{figure}

The dynamical respond of the black hole-torus system to perturbations presents  incredible evidences about
how the high-energetic phenomena occur close to the compact objects. In order to extract the information from
these phenomena and to compare them with the observation, we need to understand the dynamics of the torus-black hole 
system  in detail. In the previous subsections we have discussed the dynamical respond of this system to the high-density 
perturbation (i.e. $\rho_p = 0.9\rho_c$).  In addition to high-density perturbation we also perform a perturbative analysis
with a low density  (i.e. $\rho_p = 10^{-4}\rho_c$), which is more realistic in the astrophysical system, 
to black hole-torus system. The all numerical results from 
high-density perturbation, low-density perturbation and non-perturbation are plotted and given in Figs.
\ref{Smal_Disk_compare} and \ref{Big_Disk_compare} for models $P_3$ and $P_1$, respectively.

The initial torus given in model $P_3$ in Table \ref{table:Initial Models1} is stable and its size  is 
$r_{out}-r_{in}=11.319M$. The evolution of the rest-mass density of this torus without a perturbation is not exactly 
constant and slightly decay during the evolution due to the small amount of matter accreted onto the black hole 
(c.f. Fig.\ref{Smal_Disk_compare}). The similar trend was also found by \citet{ZRF1}.
On the other hand, the instability is triggered by perturbation and  the interaction of the size of the torus 
with a low- or high-density perturbation produces  
Papaloizou-Pringle instability,  but the time of the formation of the instability for low-density perturbation is 
slightly larger than the high-density one.  
It is apparent from Fig.\ref{Smal_Disk_compare} that the instability is fully developed
 and the highest rest-mass density of the torus reaches $\sim 50\%$ of its initial value in case 
of the low-density perturbation 
during the less than one orbital period ($t\sim 120M$ which equals to  $\sim 0.79$ orbital period) 
just after the interaction,  but it is $\sim 6\%$ in case of the high-density perturbation.
A considerable amount of matter falls  onto the black  hole  after the instability is developed and 
it oscillates during the time. And, the high-density perturbation causes the faster loss of 
matter if we compare with the low density perturbation. It is also noted that the oscillation properties of the black 
hole-torus system for both cases prominently different and they exponentially decrease over time. But the high-density 
perturbation case goes rapidly toward to zero.
The development of the non-axisymmetric modes and 
comparison of the different physical parameters in the instabilities are explained in the next 
Subsection \ref{Fourier Mode Analysis of the Torus} in detail.

Revealing the impacts of the size of the torus, location of the inner radius of the torus and 
the severity of the density of the substance used as a perturbation is important 
to understand the physical mechanism in the astrophysical phenomena. Fig.\ref{Big_Disk_compare} represents how the  
different values of the rest-mass densities of perturbations affect the bigger size of the torus with a inner radius 
located at $r_{min}=15M$. The initial torus 
given in Fig.\ref{Big_Disk_compare} has size $57.99M$, which is almost $\sim 5$ times larger than the model $P_3$ 
and it is perturbed by low- and high-rest-mass density. It is 
observed that if the density of the perturbation is larger, accretion toward to the torus has a considerable amount
and therefore a high amplitude quasi-periodic oscillation is developed very rapidly. On the other hand, if the 
magnitude of the rest-mass density of the perturbation is small, less mass accreates during the evolution. Hence more time
will be needed to produce the instability and the quasi-periodic oscillation  as shown in Fig\ref{Big_Disk_compare}. 
In this case, oscillation grows slowly for a smaller accretion rate. In addition to these,  it is also noted that the normalized 
rest-mass density in the case of the non-perturbation is almost constant.

We conclude this subsection by comparing  Figs.\ref{Smal_Disk_compare} and  \ref{Big_Disk_compare}
that the size of torus, the location of the inner radius of the torus  and the amount of the perturbation 
play an important role to determine the time of the formation of the Papaloizou-Pringle instability and the 
quasi-periodic oscillation 
in the black hole-torus system.

%

%%%%%%%%%%%%%%%%%%%%%%%%%%%%%%%%%%%%%%%%%%%%%%%%%%%%%%%%%%%%%%%%%%%%%%%
%%%%%%%%%%%%%%%%%%%%%%%%%%%%%%%%%%%%%%%%%%%%%%%%%%%%%%%%%%%%%%%%%%%%%%%

\subsection{Fourier Mode Analysis of the Instability of the Torus }
\label{Fourier Mode Analysis of the Torus}

%%%%%%%%%%%%%%%%%%%%%%%%%%%%%%%%%%%%%%%%%%%%%%%%%%%%%%%%%%%%%%%%%%%%%%%
The non-axisymmetric instabilities can be identified and characterized by using the 
linear perturbative studies of the black hole-torus system. The characterization of the 
instability, which is carried out by defining the azimuthal wave number $m$ and computing the saturation point,  
is performed from the simulation data by computing Fourier power  in density.
The Fourier modes $m=1$ and $m=2$ and growth rate are obtained at the equatorial plane (i.e. $\theta = \pi/2$) 
using the following equations given in \citet{DeVHaw},

\begin{eqnarray}
Im (w_m(r)) = \int_{0}^{2\pi} \rho(r,\phi)sin(m \phi) d\phi,
\label{Mode_Power1}
\end{eqnarray}

\begin{eqnarray}
Re (w_m(r)) = \int_{0}^{2\pi} \rho(r,\phi)cos(m \phi) d\phi,
\label{Mode_Power2}
\end{eqnarray}

\noindent
From the Eqs.\ref{Mode_Power1} and \ref{Mode_Power2}, the mode power is defined as

\begin{eqnarray}
P_m = \frac{1}{r_{out}-r_{in}}\int_{r_{in}}^{r_{out}} ln\left([ Re (w_m(r))]^2 + [Im (w_m(r))]^2\right)dr,
\label{Mode_Power3}
\end{eqnarray}

\noindent
where $r_{in}$ and $r_{out}$  are the inner and outer radii of the initially formed  stable torus. The values of these radii 
are given in Table \ref{table:Initial Models1} for all models.

\begin{figure}
 \center
\vspace{0.3cm}
\psfig{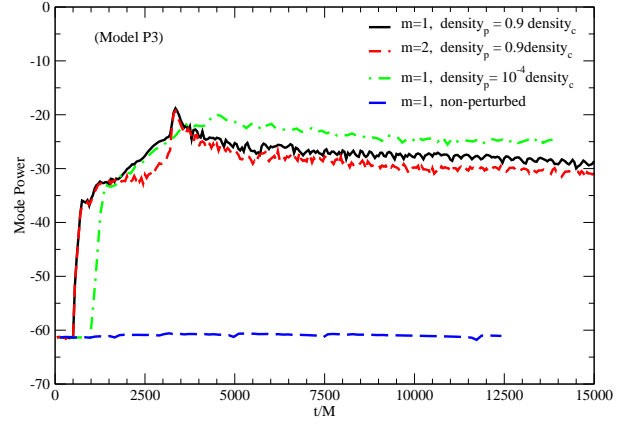}
\caption{Evolution of the modes growth for perturbed and non-perturbed torus around 
the black hole for an initial data given in model $P_3$.
\vspace{0.3cm}
\label{Small_disk Mode1}}
\end{figure}

The growth of the main Papaloizou-Pringle instability modes in case of high- and low-density perturbations, 
and without any perturbation for model $P_3$ is given in Fig.\ref{Small_disk Mode1}.  The mode powers of 
$m=1$ and $m=2$ are shown as a function of time.  It is shown that the modes grow significantly for torus. 
As expected, non-perturbed black hole-torus system does not 
present any mode growing during the evolution. After about $t=502M$ times which is equal to $t=3.3t_{orb}$ 
orbital period, the torus starts to develop $m=1$ and $m=2$ non-axisymmetric modes and these modes
undergo exponential growth until $t=757M$ strongly. Both modes grow together until 
about $t=1741M$. They diverge at this time and converge again at the saturation point.  
The saturation  point  represents  the largest 
value of the mode amplitude which is reached at  $t_{sat}=3351M$ (i.e. $t_{sat}=22t_{orb}$) 
for modes $m=1$ and $m=2$.
It is important to note that the non-axisymmetric  dynamics survives with an remarkable amplitude even after 
the saturation of the  Papaloizou-Pringle instability. Meanwhile, the $m=2$ mode shows more erratic behavior
between $t=1741M$ and $t=3351M$. 
The  $m=1$ deformation for the low-density  perturbation, $\rho_p = 10^{-4} \rho_c$, is slightly different 
than the high-density perturbation. The exponential growth in the high-density perturbation stars at later time
(i.e. $t=990M$) and reaches its peak value at $t=4542M$ which is called the saturation point.

%%%%%%$m=1$ is the fastest growing mode in all models.

%%%Low m mode is most unstable 

%

\begin{figure}
 \center
\vspace{0.3cm}
\psfig{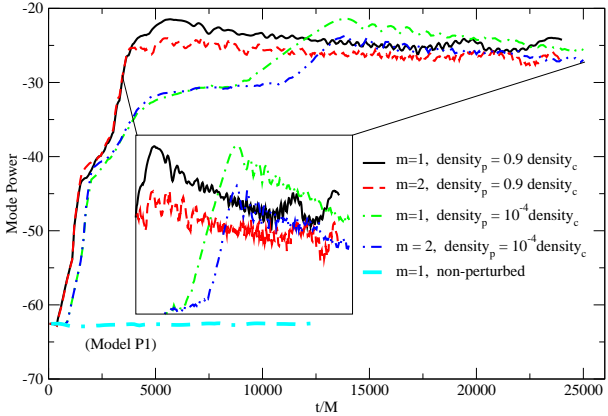}
\caption{Same as Fig.\ref{Small_disk Mode1}  but for model $P_1$.
\vspace{0.3cm}
\label{Big_disk Model1}}
\end{figure}

Unlike the model $P_3$, the amplitudes at the saturation points of $m=1$ and $m=2$ modes do not coincide 
for model $P_1$. The maximum amplitude of the $m=1$ mode is always larger,
but the saturation times for both modes  almost  equal to each other, $t_{sat}=5681M$ 
($t_{sat}=7.2t_{orb}$) for $\rho_p = 0.9 \rho_c$
and $t_{sat}=13889M$ ($t_{sat}=17.7t_{orb}$) for $\rho_p = 10^{-4} \rho_c$ as shown in Fig.\ref{Big_disk Model1}.
The $m=1$ and $m=2$ modes grow rapidly regardless of the initial values of the perturbations. But the mode in the 
high-density perturbation starts to grow early than the low-density one. After they reach to the saturation point,
the behaviors of the mode powers  almost remain the same during the rest of the simulation (c.f. the inset of 
Fig.\ref{Big_disk Model1}.) The difference between the saturation points of the high- and low-density 
perturbations is $t=8232M$ $(t = 10.5 t_{orb})$.

The mode power given in Figs.\ref{Small_disk Mode1} and \ref{Big_disk Model1} and  the mass accretion rate given in 
the left panel of the Fig.\ref{Mass_acc1} clearly show that the significant mass accretion appears when the 
growing mode of the Papaloizou-Pringle instability reaches to the saturation point. It happens due to the spiral 
wave generated as a consequence of interaction between black hole-torus system and matter used as a perturbation. The
spiral wave, and therefore,  Papaloizou-Pringle instability transports the angular momentum of the 
torus through the co-rotation radius.

\begin{figure}
 \center
\vspace{0.4cm}
\psfig{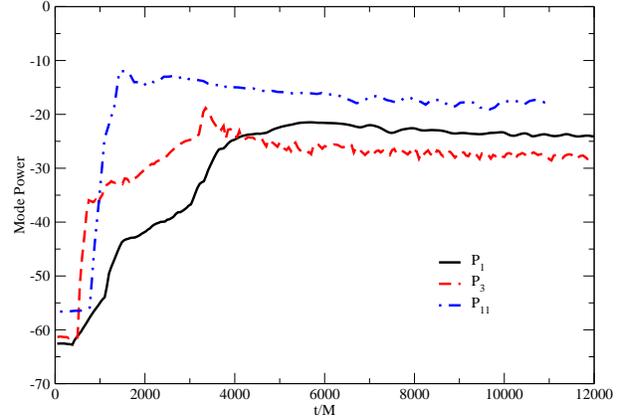}
\caption{The m=1 growing mode amplitude for different models, $P_1$, $P_3$ and $P_{11}$.
Models $P_1$, $P_3$ are for  the  different size of the torus around the non-rotating black hole but
$P_{11}$ represents the $m=1$ mode for the perturbed torus around the rotating black hole.
\vspace{0.3cm}
\label{P1_P3_P11_m1}}
\end{figure}

Mode powers of $m=1$ deformations for non-rotating black hole with a small (model $P_3$) 
and bigger  (model $P_1$)  sizes of an initial tori,
and  for a rotating black hole (model $P_{11}$) are computed and given in Fig.\ref{P1_P3_P11_m1}.
As it is displayed in Fig.\ref{P1_P3_P11_m1},  the growing times, maximum amplitude  and 
developed saturation points of the m=1 non-axisymmetric mode are modified by the size of the torus, 
the black hole spin and the density of  the perturbation.
The interaction of the non-axisymmetric 
$m=1$ deformation  with a black hole spin leads to the formation of a high-amplitude in the growth rate 
and it reaches to the saturation point. While the maximum mode power of the Papaloizou-Pringle instability is weak for 
the bigger size of the torus,  it has vigorous one for  small size of torus and  torus around the 
rotating black hole. It is hard to define the exact definition of the saturation time because of the 
weak-maximum power mode.

%%The weak-maximum power mode makes an exact definition of the saturation time difficult.}

%
%

%%%%%%%%%%%%%%%%%%%%%%%%%%%%%%%%%%%%%%%%%%%%%%%%%%%%%%%%%%%%%%%%%%%%%%%
\subsection{QPOs from the oscillating Torus}
\label{QPOs from the oscillating Torus}

%%%%%%%%%%%%%%%%%%%%%%%%%%%%%%%%%%%%%%%%%%%%%%%%%%%%%%%%%%%%%%%%%%%%%%%

\begin{figure*}
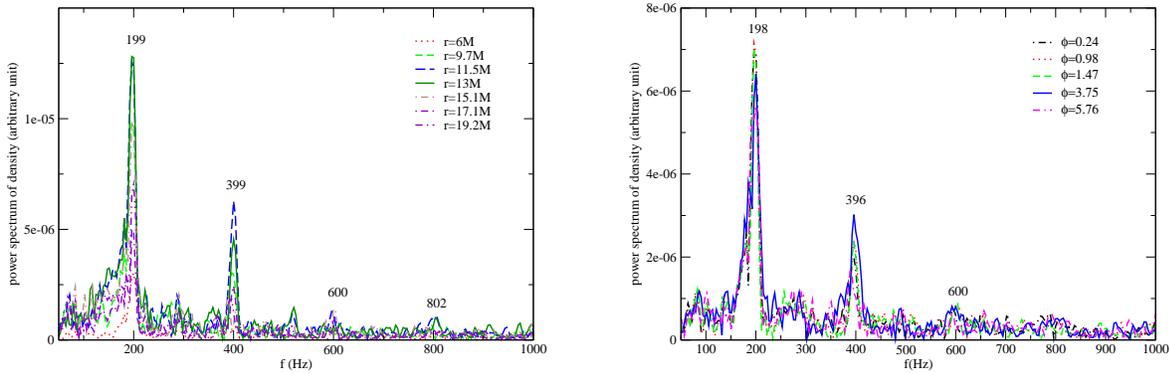

 \center
\vspace{0.5cm}
\psfig{file=Cs_001_Big_Tori_M2_G43_PS_phi_0_12_all_r.eps,width=7.2cm}
\hskip 1.0cm
\psfig{file=Cs_001_Big_Tori_M2_G43_PS_r_8_12_all_phi.eps,width=7.2cm}
\caption{Power spectrum of the rest-mass density at a single point
(any $\phi$ or $r$) for $M=2.5\,M_{\odot}$ for model  $P_1$ shown in 
Table \ref{table:Initial Models1}.
A fundamental frequency appears at far left of the both graphs. The others are 
the consequence of overtones.
The computed frequencies at any $r$ or $\phi$ overlap.}
\label{PSD_P1}
\end{figure*}

We have performed the simulations using different values of $\ell_{0}$,  $r_{cusp}$,  $r_{in}$ and $r_c$ 
of the initial torus with a perturbation given in Table \ref{table:Initial Models1}. For 
these initial perturbed discs, the global oscillations are seen at various frequencies. 
The frequencies of oscillating torus are computed by using the proper time.
The unit of the computed frequency is in the geometrized unit and it is translated to Hertz of frequency
using the following equation,

\begin{eqnarray}
f(Hz) = f(M)\times 2.03027 \times 10^5 \times \left(\frac{M_{\odot}}{M}\right),
\label{Trans_Geo_Hz}
\end{eqnarray}

\noindent
where $M$ is the mass of black hole, $f(M)$ is the frequency in the    geometrized unit, 
$M_{\odot}$ is the solar mass and $f(Hz)$  is the frequency in $Hz$. 
Applying a perturbation to a perfectly stable torus gives rise to an epicyclic motion in radial direction
due to the oscillation mode. The radial epicyclic frequency is
the frequency of the displaced matter oscillating in radial direction.
Fig.\ref{PSD_P1} consists of a fundamental frequency  at $\sim199 Hz$ and
number of overtones, which are the results of the non-linear couplings,
 $\sim 399 Hz $,  $\sim 600 Hz$ , $\sim 800 Hz$ etc. that 
they can determine the ratios 1:2:3:.... It suggests that the non-linear oscillation 
of the torus due to the perturbation is a consequence of fundamental modes of the torus.
The power law distribution given in the right panel of Fig. \ref{PSD_P1} indicates how
the peaks appear as a function of $\phi$ at fixed $r=8.12M$ which almost represents
the point of maximum density of the torus. 
The amplitudes for the corresponding peaks are almost the same for all 
$\phi$ at the same frequencies. In both graphs, two strong narrow peaks locate at 
$\sim199 Hz$ and  $\sim 399 Hz $.  
Fig. \ref{PSD_P1} also indicates that QPO frequencies are global and just confined within the 
torus.  We note from a long history of experience in our numerical models
that the computed QPO frequencies do not depend on the excision radius defined
in Boyer-Lindquist coordinate.
The QPO behavior of the oscillating torus found in our work 
might be used to explain commensurability frequencies observed in the sources, 
$H1743−322$, $XTE$ $J1550–564$, and $GRS$ $1915+105$ \citep{McCRem}.

The QPO's frequencies from our simulation for the black hole $M=2.5\,M_{\odot}$ are
approximately the same as found in \citet{ZanRez,ZRF1}. They computed the power spectrum 
of the $L_2$ norm error of the rest-mass density for different 
$\eta$ after the long time evolution. 
 The fundamental frequencies, which are seen around $200Hz$, 
and the series of overtones for both  cases are observed. 
Our study indicates 
that the perturbed torus, which 
is caused either by the introduction of a suitable parametrized perturbation to the 
radial velocity done by \citet{ZanRez,ZRF1} 
or by a matter coming from the outer boundary toward the torus (our simulations), produces 
regular oscillatory behavior.
The modes observed from both perturbations are called the pressure $(p)$ mode of oscillation of 
the torus.  
The small size of accretion torus orbiting around the black hole can be used to explain $p$ mode 
in high frequency quasi-periodic oscillations \citep{RYZ1}.
The $p$ mode represents the oscillation of matter in the horizontal direction at the equatorial
plane and is trapped inside the torus.
As shown in Figs.\ref{Model P1_2} and \ref{Mass_acc1}, 
the structure of the torus changes significantly due to the angular momentum transport of the torus. So
the $p$ mode is able to survive during the evolution.
Due to the two-dimensional structure of the system, the 
magnetic field did not create a strong effect on the wave properties of $p$ mode\citep{FuLai}.

%\noindent
It is known from previous discussion that the torus loses matter during time evolution,
and it has chaotic behavior seen in models $P_3$ and $P_4$. 
Due to this irregular non-linear oscillation, only a fundamental mode appears, but their 
overtones are absent in the power spectrum when the torus is initially more close to the black hole.  
We suggest that the variation of the torus's matter results from the non-spherically 
symmetric perturbation of the inner accreted torus. 

We have also analyzed the time evolution of mass accretion rate for the model, $P_{11}$ reported
in Table \ref{table:Initial Models1} to compute Fourier spectra of the  perturbed torus around the 
rotating black hole. After the perturbation starts to influence the torus dynamics, mass accretion 
rate oscillates around its quasi-equilibrium point. Due to a distinctive quasi-periodic 
oscillation of the unstable torus, 
the periodic character of the  spiral structure becomes more evident after perturbation reaches 
to the torus a long time later.   After the torus relaxes to regular 
oscillation, it loses the mass, and 
the oscillation amplitude decreases during the evolution. The computing QPO frequencies from 
the mass accretion rate might not give definite information, but we may approximately predict it. 
The power spectra of mass accretion rate shows a fundamental mode with a strong amplitude at 
$500Hz$ and series of overtones, as a consequence of the non-linear couplings, 
located at $1001Hz$, $1491Hz$, and $1998Hz$ for $a=0.9$ and 
$M=2.5\,M_{\odot}$. It is worth emphasizing that a fundamental frequency from the torus around the 
rotating black hole is much larger than the non-rotating case seen in Table \ref{table:QPOs}. 
This is an acceptable result
because the torus is more closer, and gravity is more stronger in case of the rotating black hole.
Table \ref{table:QPOs} also suggests that a linear scaling is presented  for the black hole spin.

The observed high frequency QPOs  around the black holes and neutron stars were described 
by \citet{KulAbram, AbramKul, BSAKT, Mukhop, MukBS} using a single model which has 
addressed the variation of QPO frequencies.
Based on the proposed model, they had predicted the spin of the black hole using the observed QPO
pairs which seem to appear at a $3:2$ ratio. The lower and higher frequencies of pairs for any system are
$\nu_l= \nu_{\perp} - \nu_s/2$ and $\nu_h= \nu_{r} + \nu_s$, respectively. 
Here, $\nu_{\perp}$, $\nu_{r}$  and 
$\nu_s$ are theoretical values of radial,  vertical epicyclic frequencies \citep{OrhanQPO} and
spin frequency of the black hole,  respectively. Using the above model we find that the computed 
commensurable frequency, $600:399$, is exposed  at $\sim 8.1M$. The estimated value of location of
resonance from numerical computation, which is $\sim 9M$, is in the suitable range of theoretical
calculation.

\begin{table*}
\scriptsize 
  \caption{The frequencies of genuine mode and its series of overtones, due to the 
nonlinear coupling, inside
the torus from the evolution of mass accretion rate. It is computed for different 
black hole spins while the mass of the black hole is assumed to be $M=2.5\,M_{\odot}$.
 \label{table:QPOs}}
\begin{center}
\vspace*{-2ex}
  \begin{tabular}{cccccc}
    \hline 
    \hline 
 Model  & $a/M$    & $f_1[Hz]$ & $o_1[Hz]$ & $o_2[Hz]$ &  $o_3[Hz]$ \\
   \hline 
 $P_1$  & $0.0$  & $199$ & $399$ & $600$ &  $800$ \\
 $P_{11}$  & $0.9$  & $500$ & $1001$ & $1491$ & $1998$ \\
\hline 
\hline 
  \end{tabular}
\end{center}
%  \tablenotetext{}{}
%\vskip -0.8truecm
\end{table*}
%

%%%%%%%%%%%%%%%%%%%%%%%%%%%%%%%%%%%%%%%%%%%%%%%%%%%%%%%%%%%%%%%%%%%%%%%
%%%%%%%%%%%%%%%%%%%%%%%%%%%%%%%%%%%%%%%%%%%%%%%%%%%%%%%%%%%%%%%%%%%%%%%
%%%%%%%%%%%%%%%%%%%%%%%%%%%%%%%%%%%%%%%%%%%%%%%%%%%%%%%%%%%%%%%%%%%%%%%
%%%%%%%%%%%%%%%%%%%%%%%%%%%%%%%%%%%%%%%%%%%%%%%%%%%%%%%%%%%%%%%%%%%%%%%

\section{Conclusion}
\label{Conclusion}
%%%%%%%%%%%%%%%%%%%%%%%%%%%%%%%%%%%%%%%%%%%%%%%%%%%%%%%%%%%%%%%%%%%%%%%
We have performed the numerical simulation of the perturbed accreted torus around the 
black hole and uncovered the oscillation properties to compute QPOs. Instead of giving 
a perturbation to the radial velocity or to the density of the stable torus 
studied by different authors in literature, 
the matter is sent from the outer boundary toward the torus-black hole system to perturb it. 
We have initially considered the different sizes of the stable tori and torus-to-hole mass 
ratios to expose oscillations of the tori, and computed the  fundamental modes and their overtones
which results from the excitation of frequencies due to perturbation.

%\noindent
We have confirmed that the torus around the black hole might have a quasi-periodic oscillation 
only if we choose a suitable initial data for a stable accretion torus. 
There are a number of different physical parameters which may affect the oscillation properties
of the torus. It is seen from our numerical simulations that the size of torus  and its initial radius 
affect the frequencies
of oscillating torus and their overtones. The size of torus and its  rest-mass density also
influence  the time of onset of the instability and
the power of oscillations and, this oscillation creates radiation of hotter photons. The similar
conclusion has been also confirmed by \citet{Bursa}. The oscillation and mass 
losing rate of the perturbed torus strongly depend on the Mach number of perturbing matter.  
Upon the given appropriate parameters of perturbation,   the subsonic or mildly supersonic flow
produces the pressure-supported oscillating torus around the
black hole and the amplitude of oscillation gradually decreases while the Mach number of 
perturbation increases. 
For the most of the models  given in Table \ref{table:Initial Models1}, 
we have found that the rest-mass density of the torus substantially 
decreases due to the Papaloizou-Pringle instability which is developed as a consequence of
interaction between the propagation of waves across the co-rotation radius.
The mode powers of 
$m=1$ and $m=2$ grow as a function of time. It is depicted that the modes grow significantly for torus. 
On the other hand, a torus which has a bigger inner radius and size,  develops a new inner
radius and cusp location after a perturbation.  During this progress, a Papaloizou-Pringle instability
is developed.
Later, the black hole-torus system reaches a new quasi-steady
state and does not present a Papaloizou-Pringle instability. 
Thus, the inner radius, the specific angular momentum of torus, size of the torus, and 
Mach number  and density of the perturbation  
play a critical role in the onset of the Papaloizou-Pringle instability.

Our studies also indicate that QPOs are common phenomena on the disc around the black holes. 
If the accretion disc or torus
have a quasi-periodic behavior, it emits continuous radiation during the computation. The amplitude
of oscillation is excited by nonlinear physical phenomena. It has been explained in terms
of the excitation of pressure gradients on the torus  and it is called $p$ mode.  It is shown that 
the power law distribution of oscillating stable torus includes a fundamental frequency  
at $\sim199 Hz$ and their number of overtones, $\sim 399 Hz $,  $\sim 600 Hz$, and $\sim 800 Hz$
that they can  determine the ratios $1:2:3:4$.
These frequencies are observed at any radial and angular directions of the torus.  The computed
QPO frequencies are almost the same as the ones given by  \citet{ZanRez,ZRF1} even though they apply 
a perturbation to the radial velocity or to the density of the stable torus 
in order to have an oscillation. 
The mode called as p is also the same in both simulations.
We have also confirmed that the fundamental
QPO frequencies and their overtones for a rotating black hole are much higher 
than the non-rotating case due to
strong gravity and location of the inner radius of the accreted torus.  
The strong gravity plays a dominant role in 
the high frequency modulation of observed X-ray flux in the binary system 
\citep{BuAbKaKL}. It is consistent 
with the computation from our simulations seen in Table \ref{table:QPOs}.

%We have also reported that the initial perturbation induces the 
%torus-black hole system to start the 
%instability around the rotating black hole before the transferred matter reaches to the black hole. The 
%perturbation can also lead the matter of the torus to fall into the black hole. It reduces 
%significantly the total rest-mass density of the torus with time, and the oscillatory behavior of the 
%rest-mass density changes depending on the inner radius of the torus. 

Finally, we have performed the numerical simulation using $2D$ code on the equatorial plane. Clearly,
to investigate the effects of forces in the vertical direction on the torus instability that
may play an important role on the dynamics of the whole system, $3D$ numerical simulations are required. The 
mode coupling in the vertical direction is likely to affect the instability of a perturbed torus-black 
hole system. Therefore, we plan to model the same perturbed system by using $3D$ code in the future.

\section*{Acknowledgments}
We would like to thank the anonymous referee for constructive comments on the original
version of the manuscript.
The numerical calculations were performed at the National Center for
High Performance Computing of Turkey (UYBHM) under grant number
10022007.


\begin{thebibliography}{} 
\bibitem[Abramowicz \& Fragile(2013)]{AbramFrag}Abramowicz, M. A. \& Fragile, P. C. 2013, Living Rev. 
Relativity, 16, arXiv:1104.5499.
\bibitem[Abramowicz et.al.(1983)]{AbrCalNob}Abramowicz, M. A., Calvani, M. \& Nobili, L. 1983, Nature, 302, 597-599.
\bibitem[Abramowicz \& Kluzniak (2001)]{AbramKul}Abramowicz, M. A., \& Kluzniak, W. 2001, A\&A, 374, L19-L20. 
\bibitem[Abramowicz et.al.(1978)]{AbrJarSik}Abramowicz, M. A., Jaroszynski, M. \& Sikora, M. 1978, 
A\&A, 63, 221-224. 
\bibitem[Blaes et.al.(2007)]{BSAKT}Blaes, O. M., Sramkova, E.,  Abramowicz, M. A., Kluzniak, W. \& 
Torkelsson, U. 2007, ApJ, 665, 642-653.
\bibitem[Blaes \& Glatzel (1986)]{BlaGlat}Blaes, O. M. \& Glatzel W. 1986, MNRAS, 220, 253.
\bibitem[Burkert et.al. (2012)]{BSAGGFEE}Burkert, A., Schartmann, M., Alig, C., Gillessen, S., Genzel, R., Fritz, T. K.
\&  Eisenhauer, F. 2012, ApJ, 750, 58B.
\bibitem[Bursa (2005)]{Bursa}Bursa, M. 2005, Astron.Nachr. 326, 849-855.
\bibitem[Bursa et.al. (2004)]{BuAbKaKL}	Bursa, M., Abramowicz, M. A., Karas, V. \& Kluzniak, W. 2004,
ApJ, 617, L45-L48.
\bibitem[Chen \& Beloborodov (2007)]{ChBe}Chen, W. \& Beloborodov, A. M. 2007, ApJ, 657, 383.
\bibitem[Coward et.al.(2002)]{CowPutBur}Coward, D. M., van Putten, M. H. P. M., \& Burman, R., 
R. 2002, ApJ, 580, 1024.
\bibitem[Daigne \& Font(2004)]{DaiFont}Daigne, F. \& Font, J. A. 2004, MNRAS, 349, 841.
\bibitem[De Villiers \& Hawley (2002)]{DeVHaw}De Villiers, J.-P. \& Hawley, J. 2002, ApJ, 577, 866. 
\bibitem[D\"onmez et.al.(2011)]{DZR}D\"onmez O., Zanotti, O., \& Rezzolla, L. 2011, MNRAS, 412, 1659.
\bibitem[D\"onmez (2004)]{Orhan}D\"onmez O. 2004, Astrophys. Spac. Sci., 293, 323.
\bibitem[D\"onmez (2007)]{OrhanQPO}Donmez O. 2007, MPLA, 22, 141-157. 
\bibitem[Font \& Daigne(2002)]{FonDai}Font, J. A. \& Daigne, F. 2002, MNRAS, 334, 383.
\bibitem[Font et.al.(2000)]{FonMiSuTo}Font, J.A., Miller, M., Suen, W.-M. \& Tobias, M. 2000, Phys. 
Rev. D, 61, 044011.
\bibitem[Fu \& Lai (2009)]{FuLai}Fu, W. \& Lai, D. 2009, ApJ, 690, 1386.
\bibitem[Hawley (1991)]{Hawley}Hawley J. F. 1991, ApJ, 381, 496.
\bibitem[Kluzniak \& Abramowicz (2001)]{KulAbram}Kluzniak, W. \& Abramowicz, M. A. 2001, astro-ph/0105057.
\bibitem[Korobkin et.al.(2013)]{KASSZRO}Korobkin, O., Abdikamalov, E., Stergioulas, N.,
 Schnetter, E., Zink, B., Rosswog, S. \& Ott, C. 2013, MNRAS, 431, 349-354.
\bibitem[Kiuchi et.al.(2011)]{KiShMonFo}Kiuchi, K., Shibata, M., Montero, P. \& Font, J., A. 
2011, Physical Review Letters, 106, 251102
\bibitem[Lee (2004)]{lee1}Lee, H., K. 2004, Journal of Korean Physical Society, 45, 1746. 	
\bibitem[Levinson \& Blandford (1996)]{LevBlan} Levinson, A. \& Blandford, R. 1996, ApJ, 456L, 29L.
\bibitem[McClintock \& Remillard (2004)]{McCRem}McClintock J. E. \& Remillard R. 2004, in LewinW. H. G., 
van der Klis M., eds, Compact Stellar X-Ray Sources. Cambridge University Press, Cambridge. 
\bibitem[Meszaros (2006)]{Meszaros1}Meszaros, P. 2006, Rep. Prog. Phys., 69, 2259.
\bibitem[Montero et.al.(2010)]{MFS}Montero, P. J., Font, J. A.  \& Shibata, M. 2010,  PRL, 104, 191101.
\bibitem[Montero et.al.(2007)]{MZFR}Montero, P. J., Zanotti, O., Font, J. A. \&  Rezzolla, L 2007, MNRAS, 378, 1101.
\bibitem[Moscibrodzka et.al.(2006)]{MosDasCz}Moscibrodzka, M., Das, T. K. and Czerny, B. 2006, MNRAS, 370, 219. 
\bibitem[Mukhopadhyay (2009)]{Mukhop}Mukhopadhyay, B. 2009, ApJ, 694, 387-395.
\bibitem[Mukhopadhyay et.al.(2012)]{MukBS}Mukhopadhyay, B., Bhattacharya, D. \& Sreekumar, P. 2012, 
IJMPD, 21, 1250086.
\bibitem[Nagar et.al.(2005)]{NFZde}Nagar, A., Font, J. A., Zanotti, O. \&  de Pietri, R. 2005, PRD, 72, 024007. 
\bibitem[Papaloizou \& Pringle (1984)]{PapPri1}Papaloizou, J. C. M. \& Pringle, J. E. 1984, MNRAS, 208 721.
\bibitem[Papaloizou \& Pringle (1985)]{PapPri2}Papaloizou, J. C. M. \& Pringle, J. E. 1985, MNRAS, 213 799.
\bibitem[Rees \& Meszaros (1994)]{ReMes}Rees M. J. \&  Meszaros P. 1994, ApJL, 430, L93.
\bibitem[Remillard et.al.(1999)]{RMEMJBOJ}Remillard, R. A., Morgan, E. H., McClintock, J. E.,
 Bailyn, C. D. \& Orosz, J. A. 1999, ApJ, 522, 397-412.
\bibitem[Rezzolla et.al.(2003)]{RYZ1}Rezzolla L, Yoshida S and Zanotti O 2003a, MNRAS, 344, 978.
\bibitem[Schnittman \& Krolik (2008)]{ScKr}Schnittman, J. D. \& Krolik, J. H. 2008, 684, 835.
\bibitem[Schnittman \& Rezzolla(2006)]{ScRe}Schnittman, J  \& Rezzolla, L.  2006, ApJ, 637L, 113S.
\bibitem[Strohmayer (2001)]{Strohmayer1}Strohmayer, T. E. 2001, ApJ, 552, L49.
\bibitem[van der Klis(2004)]{VanDer}van der Klis M 2004, eprint arXiv:astro-ph/0410551.
\bibitem[Fu \& Lai (2011)]{FuLai}Fu, W., \& Lai, D. 2011, MNRAS, 410, 1617.
\bibitem[Yildiran \& D\"onmez(2010)]{DenOrh} Yildiran, D.  \& D\"onmez O. 2010, IJMPD, 19, 2111-2133.
\bibitem[Zanotti et.al.(2003)]{ZRF1}Zanotti O, Rezzolla L and Font J A 2003, MNRAS, 341, 832.
\bibitem[Zanotti et.al.(2005)]{ZFRM}Zanotti, O., Font, J. A., Rezzolla, L.  \& Montero, P. J.  2005, MNRAS, 356, 1371.
\bibitem[Zanotti \&  Rezzolla(2003)]{ZanRez}Zanotti, O  \& Rezzolla, L. 2003 MSAIS, 1, 192.
\bibitem[Zurek \& Benz(1986)]{ZurBen}Zurek, W., H. \&  Benz, W. 1986, ApJ 308, 123.







\end{thebibliography}
\end{document}